\documentclass[%
 reprint,
 amsmath,amssymb,
 aps,
prb,
longbibliography,
superscriptaddress,
]{revtex4-2}

\usepackage{graphicx}
\usepackage{dcolumn}
\usepackage{bm}
\usepackage{mathrsfs}
\usepackage{bm}
\usepackage[normalem]{ulem}
\usepackage[dvipsnames]{xcolor}
\usepackage[english]{babel}
\usepackage[colorlinks=true,linkcolor=blue,citecolor=blue,urlcolor=blue,bookmarks=False]{hyperref} 

\usepackage{braket}
\usepackage{pdfpages} 
\usepackage{pgffor} 

\makeatletter
\AtBeginDocument{\let\LS@rot\@undefined}
\makeatother

\newif\ifarXiv
\arXivfalse

\newcommand{\ismpoli}{Istituto di Struttura della Materia-CNR (ISM-CNR)  and European Theoretical Spectroscopy Facility (ETSF), Piazza Leonardo da Vinci 32, 20133 Milano, Italy}
\newcommand{\nano}{Centro S3, CNR–Istituto Nanoscienze, 41125 Modena, Italy}

\newcommand{\editor}[2]{%
  \expandafter\newcommand\csname #1note\endcsname[1]{%
    \textcolor{#2}{(\textbf{#1:} \it ##1)}}%
  \expandafter\newcommand\csname #1\endcsname[1]{%
    \textcolor{#2}{##1}}%
  \expandafter\newcommand\csname #1cancel\endcsname[1]{%
    \textcolor{#2}{\sout{##1}}}%
  \expandafter\newcommand\csname #1change\endcsname[2]{%
    \textcolor{#2}{\sout{##1} ##2}}%
  \newenvironment{#1text}{\color{#2}}{\color{black}}
}

\def\replytoreferee#1{#1}

\begin{document}

\preprint{APS/123-QED}

\title{Symmetries of Excitons}

\author{Muralidhar Nalabothula}
\email{muralidharrsvm7@gmail.com}
\affiliation{Department of Physics and Materials Science, University of Luxembourg, L-1511 Luxembourg, Luxembourg.}
\author{Davide Sangalli}
\affiliation{\ismpoli}
\author{Fulvio Paleari}
\affiliation{\nano}
\author{Sven Reichardt}%
\affiliation{Department of Physics and Materials Science, University of Luxembourg, L-1511 Luxembourg, Luxembourg.}
\author{Ludger Wirtz}%
\affiliation{Department of Physics and Materials Science, University of Luxembourg, L-1511 Luxembourg, Luxembourg.}

\date{\today}
\begin{abstract}
Excitons, bound electron–hole pairs, are responsible for strong optical resonances near the bandgap in low-dimensional materials and wide-bandgap insulators. Although current \textit{ab initio} methods can accurately determine exciton energies and eigenstates, their symmetries have been much less explored. In this work, we employ standard group-theory methods to analyse the transformation properties of excitonic states, obtained by solving the Bethe–Salpeter equation, under crystal symmetry operations. We develop an approach to assign irreducible-representation labels to excitonic states, providing a state-of-the-art framework for analysing their symmetries and selection rules (including, for example, the case of exciton-phonon coupling). Complementary to the symmetry classification, we introduce the concept of total crystal angular momentum for excitons in the presence of rotational symmetries, allowing the derivation of conservation laws.
Furthermore, we demonstrate how these symmetry properties can be exploited to greatly enhance the computational efficiency of exciton calculations with the Bethe–Salpeter equation. We apply our methodology to three prototypical systems to understand the role of symmetries in different contexts: (i) For LiF, we present the symmetry analysis of the entire excitonic dispersion and examine the selection rules for optical absorption. (ii) In the calculation of resonant Raman spectra of monolayer MoSe$_2$, we demonstrate how the conservation of total crystal angular momentum governs exciton–phonon interactions, leading to the observed resonant enhancement. (iii) In bulk hBN, we analyze the role of symmetries for the coupling of finite-momentum excitons to finite-momentum phonons and their manifestation in the phonon-assisted luminescence spectra. This work establishes a general and robust framework for understanding the symmetry properties of excitons in crystals, providing a foundation for future studies.
\end{abstract}

\maketitle
\section{Introduction}
In a crystal, exciting an electron from the occupied valence band to the unoccupied conduction band creates a positively charged electron-void known as a hole. The electrostatic interaction between the electron and the hole forms a hydrogen-like bound state called an exciton~\cite{PhysRev.37.17,PhysRev.52.191}. 
In two-dimensional (2D) materials~\cite{RevModPhys.90.021001,PhysRevB.88.045412} and layered or wide-bandgap materials, such as hexagonal boron nitride (hBN)~\cite{Watanabe2004,Paleari_2018}, excitons often dominate the optical response due to reduced dielectric screening. They play a central role in optical processes such as optical absorption~\cite{He2014Jul,PhysRevB.88.045412}, resonant Raman scattering~\cite{GOLASA201453,doi:10.1126/sciadv.abb5915,Nalabothula2025Apr,McDonnell_2020}, and phonon-assisted luminescence~\cite{Cassabois2016,PhysRevLett.122.187401,Zanfrognini2023Nov,PhysRevB.99.081109}, as well as correlation-driven phenomena such as exciton condensation~\cite{Moon2025Jul} and out-of-equilibrium carrier dynamics~\cite{Li2019Apr}. Therefore, understanding excitonic properties is crucial for exploring light–matter interactions in these materials.

A key aspect of excitons is their symmetry properties, which determine their optical selection rules and interactions with other quasiparticles. Typically, exciton selection rules are explained using a hydrogenic model. For example, in materials with strong dielectric screening, excitons composed primarily of transitions at the band extrema are optically active only if they are ``s-like''~\cite{PhysRev.108.1384}. These s-like excitons involve vertical transitions in a nearly spherically symmetric k-point region around the band extrema in the Brillouin zone.

While the hydrogenic model adequately describes dipole selection rules for conventional bulk semiconductors with Wannier–Mott~ excitons~\cite{PhysRev.52.191}, it fails for excitons that deviate from this picture. For instance, excitons in monolayer transition metal dichalcogenides (TMDCs) significantly deviate from the hydrogenic Rydberg series~\cite{PhysRevLett.113.076802,PhysRevB.93.235435,PhysRevB.88.045412}, and excitons in wide-gap insulators such as LiF are Frenkel-like and delocalized in reciprocal space, unlike Wannier excitons. Furthermore, the hydrogenic model does not capture selection rules for processes such as exciton–phonon scattering. This motivates the need for more robust approaches to fully understand exciton symmetries across different materials.

\replytoreferee{Meanwhile, in finite molecular systems, the symmetry analysis of neutral electronic excitations is well established in quantum chemistry, where excited states, including electron–hole excitations, are classified according to the irreducible representations of the molecular point group~\cite{Cotton1991Jan}. In practice, this classification is obtained by analyzing the transformation properties of many-body wavefunctions under point-group operations, enabling the assignment of symmetry labels and the derivation of selection rules~\cite{Levine2014}. For the calculation of excited states, e.g. via the method of configuration interaction, quantum chemistry codes typically use symmetry-adapted configuration state functions as a basis~\cite{szabo1996modern,helgaker2000molecular}.}

\replytoreferee{While this framework is well established for finite systems, its direct extension to periodic systems is nontrivial. In crystals, excitons are expressed as linear combinations of interband transitions at different electron and hole crystal momenta. Moreover, translational symmetry, together with possible non-symmorphic operations, introduces qualitatively new features, including exciton dispersion and projective representations, which do not arise in molecular point-group symmetry. Consequently, a purely molecular point-group analysis cannot be directly transferred to momentum-resolved excitons in crystals.}

\replytoreferee{
In this work, we formulate a general symmetry framework for excitons by exploiting the invariance of the Bethe–Salpeter equation (BSE)~\cite{PhysRevB.21.4656,PhysRevB.62.4927,PhysRevLett.81.2312,PhysRevLett.80.4510} under space-group operations. Since the BSE is widely used to describe excitonic effects in both molecules~\cite{Bruneval2015Jun,Blase2018Feb,Blase2020Sep} and solids~\cite{RevModPhys.74.601,PhysRevLett.81.2312,PhysRevLett.80.4510,Martin_Reining_Ceperley_2016}, this approach provides a unified and systematically applicable route to symmetry classification in both finite and periodic systems. We demonstrate that excitons at finite crystal momentum $\mathbf{Q}$ transform according to projective representations of the little point group of $\mathbf{Q}$, in direct analogy with phonons~\cite{RevModPhys.40.1}. The resulting formalism provides a practical procedure for assigning irreducible-representation labels to excitonic states without requiring case-by-case inspection of real-space wavefunctions~\cite{PhysRevB.94.125303}. The symmetry properties of excitons thus emerge directly from the underlying many-body Hamiltonian, yielding a robust and computationally practical framework for \textit{ab initio} BSE calculations.}

Beyond symmetry classification, we introduce the concept of total crystal angular momentum for excitons. This concept allows us to derive conservation laws in the presence of rotational symmetry, analogous to the $z$ component of angular momentum in the hydrogen atom. These conservation laws naturally lead to the notion of chirality for excitons in the absence of inversion symmetry.

We further demonstrate how crystal symmetries can be exploited to improve the computational efficiency of BSE calculations. Specifically, we show that the excitonic wavefunctions over the entire Brillouin zone can be reconstructed from those in the irreducible Brillouin zone by applying the corresponding crystal symmetry operations, thereby avoiding explicit evaluation at each $\mathbf{Q}$-point. In addition, we provide explicit expressions for the projection operators that enable block diagonalization of the BSE Hamiltonian, thereby accelerating the diagonalization of the effective two-particle Hamiltonian. Furthermore, the construction of the BSE Hamiltonian itself can be significantly optimized by evaluating only a subset of matrix elements and generating the remaining ones through symmetry operations.

Finally, we demonstrate our methodology on  three prototypical systems. First, lithium fluoride (LiF), a cubic material with full octahedral point-group symmetry that hosts Frenkel-type excitons~\cite{PhysRevLett.81.2312}, where we classify excitonic dispersion curves (bands) and elucidate the role of symmetries in the optical absorption spectrum. Second, monolayer molybdenum diselenide (MoSe$_2$), which exhibits strong spin–orbit coupling and deviations from the hydrogenic Rydberg series~\cite{PhysRevB.103.155152,PhysRevLett.113.076802,PhysRevB.93.235435}, where we analyze exciton symmetries at the Brillouin-zone center and show that conservation of total crystal angular momentum governs exciton–phonon interactions, leading to resonant enhancement of the $A'_1$ mode but not of the $E'$ mode in the Raman spectrum. Third, bulk hexagonal boron nitride ($h$BN), which possesses non-symmorphic symmetries and hosts hybrid Frenkel–Wannier excitons~\cite{PhysRevLett.96.126104,PhysRevB.94.125303,Paleari_2018}, where we examine the symmetries of finite-momentum excitons and their role in exciton–phonon matrix elements that become manifest as phonon replicas in the luminescence spectrum.

Overall, our results establish a comprehensive framework for studying exciton symmetries, laying the foundation for understanding their role in optical scattering processes across a wide range of two- and three-dimensional crystals.

\section{Symmetries of electronic states} 
We start by reviewing the symmetry properties of the electronic Hamiltonian while establishing the notation.  The set of all symmetries that leave the crystal invariant forms a space group $\mathcal{G}$. The unitary operators $\hat{U}(g)$, for all $g \in \mathcal{G}$, furnish a projective representation (or a linear representation if spin is neglected) of $\mathcal{G}$ acting on the Hilbert space of electronic states~\cite{wigner59,weinbergQFTvol1,altmann2005rotations}. The electronic Hamiltonian $\mathcal{\hat{H}}$ remains invariant under the action of $g$ if and only if $\hat{U}(g)$ commutes with $\mathcal{\hat{H}}$, which is written as $[\mathcal{\hat{H}}, \hat{U}(g)] = 0$~\cite{tung1985group}.

Within the group $\mathcal{G}$, the set of all pure translational symmetries forms a normal subgroup $\mathcal{T}$ of $\mathcal{G}$. The group $\mathcal{T}$ is abelian. That is, for all $g_1, g_2 \in \mathcal{T}$,  
$g_1 \cdot g_2 = g_2 \cdot g_1$, where $\cdot$ represents the group multiplication. This implies that the space group $\mathcal{G}$ can be decomposed into disjoint left (or right) cosets of the subgroup $\mathcal{T}$ in $\mathcal{G}$, which is written as  
\begin{equation}
    \mathcal{G} \equiv \bigcup_{i=1}^{n} g_{i}\mathcal{T},
    \label{eq:Gcosetdecom}
\end{equation}
where $g_i$ are the coset representatives, $g_i\mathcal{T}$ represents the cosets of $\mathcal{T}$ in $\mathcal{G}$, and $n$ is the index of $\mathcal{T}$ in $\mathcal{G}$. Since $\mathcal{T}$ is a normal subgroup of $\mathcal{G}$, it follows that the set of all left (or right) cosets of $\mathcal{T}$ in $\mathcal{G}$ forms the quotient group, represented by $\mathcal{G}/\mathcal{T}$ which is isomorphic to the point group $\mathcal{P}$ of the crystal~\cite{el2008symmetry}. The point group $\mathcal{P}$ is the set obtained by removing the translational components from the elements of the space group $\mathcal{G}$, which may not be a subgroup of $\mathcal{G}$.

Since the Hamiltonian $\mathcal{\hat{H}}$ commutes with all elements in $\mathcal{T}$, it follows that both $\mathcal{\hat{H}}$ and $\hat{U}(g)$ for all $g \in \mathcal{T}$ can be simultaneously block-diagonalized, with each block classified according to the one-dimensional irreducible representations of $\mathcal{T}$. Upon imposing the Born–von Karman boundary condition~\cite{ashcroft1976solid}, these irreducible representations are labeled by the wavevector $\mathbf{k} = \frac{i}{N_x}\mathbf{b}_1 + \frac{j}{N_y}\mathbf{b}_2 + \frac{k}{N_z}\mathbf{b}_3$, where $N_x \times N_y \times N_z$ is the Born–von Karman supercell, $\mathbf{b}_1$, $\mathbf{b}_2$, and $\mathbf{b}_3$ are reciprocal lattice vectors, and $i \in \{0,1,\ldots,N_x{-}1\}$, $j \in \{0,1,\ldots,N_y{-}1\}$, $k \in \{0,1,\ldots,N_z{-}1\}$. This allows us to express the eigenstates of $\mathcal{\hat{H}}$ as Bloch states, which are written as~\cite{tung1985group,dresselhaus2007group,el2008symmetry}:
\begin{equation}
    \left ( \phi_{\mathbf{k},m} \right )_\sigma(\mathbf{r}) \equiv
    \phi_{\mathbf{k},m,\sigma}(\mathbf{r}) = e^{i\mathbf{k} \cdot \mathbf{r}} \left (u_{\mathbf{k},m}\right )_\sigma(\mathbf{r}),
    \label{eq:bloch_theorem}
\end{equation}
where $\left(u_{\mathbf{k},m}\right)_\sigma(\mathbf{r})$ is the periodic part of the wavefunction, determined by the structure of the potential. Here, $m$ and $\sigma$ denote the band index and spinor component, respectively. \replytoreferee{We note that when the spin index in the Bloch states is suppressed, $\phi_{k,m}(\mathbf r)$ denotes the full spinorial wavefunction, and throughout this paper its spin components are denoted interchangeably as $(\phi_{k,m})_{\sigma}(\mathbf r) \equiv \phi_{k,m,\sigma}(\mathbf r)$.
}

The set of all elements in $\mathcal{G}$ that leave the wavevector $\mathbf{k}$ unchanged (up to a reciprocal lattice vector) forms the little group $\mathcal{G}_{\mathbf{k}}$, which is a subgroup of $\mathcal{G}$. The quotient group $\mathcal{G}_{\mathbf{k}}/\mathcal{T}$ is isomorphic to the little point group $\mathcal{P}_{\mathbf{k}}$ of $\mathbf{k}$, which is a subgroup of the point group of the crystal, $\mathcal{P}$.

Now, consider a spatial symmetry $g = \{R \mid \bm{\tau}\} \in \mathcal{G}$, which transforms the position vector $\mathbf{r}$ as $\mathbf{r} \to R\mathbf{r} + \bm{\tau}$, where $R$ is an orthogonal matrix and $\bm{\tau}$ is a translation, which may also be a fraction of the lattice vectors. The action of symmetry operation $g$ on the Bloch state $\phi_{\mathbf{k},m}(\mathbf{r})$ is~\cite{altmann2005rotations,el2008symmetry}
\begin{equation}
    \left( \{R \mid \bm{\tau}\}  \phi_{\mathbf{k},m} \right)_{\sigma} (\mathbf{r}) = \sum_{\sigma'} S_{\sigma\sigma'}(R) \left(\phi_{\mathbf{k},m}\right)_{\sigma'}(R^{-1}(\mathbf{r} - \bm{\tau})),
    \label{eq:sym_action}
\end{equation}
where $S(R) = e^{-i\frac{\alpha}{2} \mathbf{\hat{n}} \cdot \vec{\sigma}}$ is a $2 \times 2$ unitary matrix acting on the spinorial part of the electronic wavefunction. Here, $\vec{\sigma} = (\sigma_x, \sigma_y, \sigma_z)$ denotes the Pauli matrices, and $\mathbf{\hat{n}}$ and $\alpha$ are the axis and angle of the orthogonal matrix $R$ (we treat improper rotations as a product of a proper rotation and inversion. As angular momentum is invariant under inversion, for improper rotations, $\mathbf{\hat{n}}$ and $\alpha$ correspond to the $-R$ matrix). If spin is neglected, $S(R)$ reduces to the $1 \times 1$ identity matrix.

Under the action of $g$, the transformed wavefunction $\left( \{R \mid \bm{\tau}\}  \phi_{\mathbf{k},m} \right) (\mathbf{r})$ remains an eigenstate of $\mathcal{\hat{H}}$, but now corresponds to a state with wavevector $\mathbf{k}' = R\mathbf{k} + \mathbf{G}$, where $\mathbf{G}$ is a reciprocal lattice vector that maps $R\mathbf{k}$ back into the first Brillouin zone~\cite{dresselhaus2007group}. Throughout this work, we adopt the convention $\phi_{\mathbf{k},m}(\mathbf{r}) = \phi_{\mathbf{k} + \mathbf{G},m}(\mathbf{r})$, known as the \emph{periodic gauge}.

Assuming that the wave functions $\phi_{\mathbf{k},m}$ form an orthonormal set (for all $\mathbf{k}$), the wavefunctions $\left( \{R \mid \bm{\tau}\}  \phi_{\mathbf{k},m} \right) (\mathbf{r})$ and $\phi_{R\mathbf{k},m}(\mathbf{r})$ represent the same physical state, differing only by a phase (or a unitary rotation for degenerate states). This implies we can express $\left( \{R \mid \bm{\tau}\}  \phi_{\mathbf{k},m} \right) (\mathbf{r})$ as:
\begin{equation}
    \left( \{R \mid \bm{\tau}\}  \phi_{\mathbf{k},m} \right) (\mathbf{r}) = \sum_{m'} \mathcal{D}_{\mathbf{k},m'm}(g) \phi_{R\mathbf{k},m'}(\mathbf{r}),
    \label{eq:intro_dmat}
\end{equation}
where $\mathcal{D}_{\mathbf{k}}(g) = \bigoplus_i \mathcal{D}^i_{\mathbf{k}}(g)$ is a block-diagonal unitary matrix, with each block $\mathcal{D}^i_{\mathbf{k}}(g)$ corresponding to a degenerate subspace that is preserved under the action of all symmetry operations. The matrix $\mathcal{D}_{\mathbf{k}}(g)$ is given by:
{
\begin{equation}
\begin{aligned}
    \mathcal{D}^{\vphantom{*}}_{\mathbf{k},m'm}(g) = \sum_{\sigma,\sigma'} \int \phi^*_{R\mathbf{k},m',\sigma'}(\mathbf{r}) S^{\vphantom{*}}_{\sigma'\sigma}(R) &\\
    \phi^{\vphantom{*}}_{\mathbf{k},m,\sigma}(R^{-1} (\mathbf{r} - \mathbf{\bm{\tau}})) \, d^3\mathbf{r}.
    \label{eq:dmats}
\end{aligned}
\end{equation}}
The unitary matrices $\mathcal{D}_{\mathbf{k}}(g)$ for all $g \in \mathcal{G}$ and $\mathbf{k}$-points are central to the analysis of electronic/excitonic state symmetries in crystals. If $g$ belongs to the little group of $\mathbf{k}$, then $\mathcal{D}_{\mathbf{k}}(g)$ corresponds to the representation matrices of the symmetry operation in the single-particle Bloch basis, which can be further decomposed into irreducible representations~\cite{Iraola2022Mar,Matsugatani2021Jul}~\footnote{In the case of accidental degeneracies, each block $\mathcal{D}_{\mathbf{k},i}(g)$ can be further decomposed; otherwise, each block constitutes an irreducible representation.}.

We can extend $\mathcal{D}_{\mathbf{k}}$ matrices to incorporate time-reversal symmetry $\mathscr{T}$, which are given by  
\begin{equation}
\begin{aligned}
    \mathcal{D}^{\vphantom{*}}_{\mathbf{k},m'm}(\mathscr{T}) = \sum_{\sigma,\sigma'} \int \phi^*_{-\mathbf{k},m',\sigma'}(\mathbf{r}) S^{\vphantom{*}}_{\sigma'\sigma}(\mathscr{T}) &\\
    \phi^*_{\mathbf{k},m,\sigma}(\mathbf{r}) \, d^3\mathbf{r}.
    \label{eq:dmats_trev}
\end{aligned}
\end{equation}
where $S(\mathscr{T}) = -i\sigma_y$, or the $1 \times 1$ identity matrix if spin is neglected.

\section{Symmetries of excitonic states} Building on this foundation, we now examine the symmetries of excitonic states. The exciton energies and eigenstates are obtained by solving the BSE~\cite{PhysRevB.62.4927,PhysRevLett.81.2312,PhysRevLett.80.4510}. Therefore, we look at the symmetries of the BSE to understand the symmetry properties of excitons. Instead of directly working with the BSE, we first present the symmetry analysis for an arbitrary four-point function that is invariant under symmetry operations. This approach is then applied to the BSE as a special case. 

\begin{figure}
    \centering
    \includegraphics[width=0.9\linewidth]{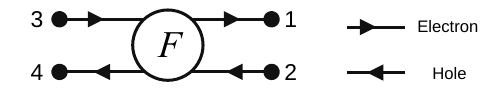}
    \caption{Feynman diagram for an arbitrary four-point function $F(1,2;3,4)$.}
    \label{fig:Fpt_feydiag}
\end{figure}

Consider a generic four-point function $F(1,2;3,4)$, represented by the Feynman diagram shown in Fig.~\ref{fig:Fpt_feydiag}, where the indices denote position, spin, and time coordinates, e.g., $1 \to (\mathbf{r}_1 \sigma_1, t_1)$. The Fourier transform of $F(1,2;3,4)$ in the time domain, denoted as $\tilde{F} = \tilde{F}(\mathbf{r}_1 \sigma_1, \mathbf{r}_2 \sigma_2; \mathbf{r}_3\sigma_3, \mathbf{r}_4\sigma_4; \omega, \omega', \omega'')$, can be expanded in terms of non-interacting one-electron Bloch states $\phi(\mathbf{r})$, which form an orthonormal basis, and is given by~\cite{sven_thesis,Stefanucci2013}
\footnote{Due to time-translation symmetry, the number of independent frequency variables in the Fourier transform is reduced from four to three~\cite{Stefanucci2013}.}:
\begin{equation}
\begin{aligned}
    \tilde{F} = \sum_{\substack{ {\mathbf{k}_1, \mathbf{k}_2, \mathbf{k}_3, \mathbf{k}_4}\\ {a, b, c, d}}} &
    \phi^{\vphantom{*}}_{\mathbf{k}_1, a ,\sigma_1}(\mathbf{r}_1) 
    \phi^{\vphantom{*}}_{\mathbf{k}_4, d, \sigma_4}(\mathbf{r}_4) 
    \phi^*_{\mathbf{k}_2, b, \sigma_2}(\mathbf{r}_2) \\ &\times 
    \phi^*_{\mathbf{k}_3, c, \sigma_3}(\mathbf{r}_3) 
    \tilde{F}_{\substack{ {\mathbf{k}_1a, \mathbf{k}_2 b} \\ {\mathbf{k}_3c, \mathbf{k}_4d}}} (\omega, \omega', \omega'') ,
\end{aligned}
\label{eq:four_pt_fun_ksbasis}
\end{equation}
where the matrix $\tilde{F}_{\substack{ {\mathbf{k}_1a, \mathbf{k}_2 b} \\ {\mathbf{k}_3c, \mathbf{k}_4d}}} (\omega, \omega', \omega'')$ represents the Fourier transform $\tilde{F}$ in the basis of one-electron Bloch states. Here, the indices $a$ and $c$ denote the band indices of the electron, while $b$ and $d$ correspond to the band indices of the hole. 

The action of $g$ on $\tilde{F}$ is given by 
\begin{equation}
    \{R \mid \bm{\tau}\} \tilde{F}  = \tilde{F}' = \tilde{F}(\mathbf{r}'_1 \sigma'_1, \mathbf{r}'_2 \sigma'_2; \mathbf{r}'_3\sigma'_3, \mathbf{r}'_4\sigma'_4; \omega, \omega', \omega''),
    \label{eq:fpt_g_action}
\end{equation}
where $\mathbf{r}_i' = R^{-1}(\mathbf{r}_i-\bm{\tau}) \ \forall i \in \{1,2,3,4\}$. Combining Eqs.~\eqref{eq:four_pt_fun_ksbasis} and \eqref{eq:fpt_g_action}, and employing Eq.~\eqref{eq:intro_dmat} we obtain

\begin{equation}
\begin{aligned}
    \tilde{F}' =  \sum_{\substack{ { {\mathbf{k}_1, \mathbf{k}_2, \mathbf{k}_3, \mathbf{k}_4}}\\ {a', b', c', d'}}} &
    \phi^{\vphantom{*}}_{R\mathbf{k}_1, a',\sigma'_1}(\mathbf{r}_1) 
    \phi^{\vphantom{*}}_{R\mathbf{k}_4, d',\sigma'_4}(\mathbf{r}_4)
    \phi^*_{R\mathbf{k}_2, b',\sigma'_2}(\mathbf{r}_2) 
    \\
    &\times
    \phi^*_{R\mathbf{k}_3, c',\sigma'_3}(\mathbf{r}_3) 
     \sum_{a, b, c, d}\Big\{ \ \tilde{F}_{\substack{ {\mathbf{k}_1a, \mathbf{k}_2 b} \\ {\mathbf{k}_3c, \mathbf{k}_4d}}} (\omega, \omega', \omega'') \\&
    \mathcal{D}^{\vphantom{*}}_{\mathbf{k}_1,a'a}(g) \mathcal{D}^*_{\mathbf{k}_3,c'c}(g) \mathcal{D}^*_{\mathbf{k}_2,b'b}(g) \mathcal{D}^{\vphantom{*}}_{\mathbf{k}_4,d'd}(g) \Big \}.
\end{aligned}
\label{eq:four_pt_fun_ksbasis_rot}
\end{equation}

If $\tilde{F}$ is invariant under the action of $g$, i.e  $\tilde{F} = \tilde{F}'$ then from Eq.~\eqref{eq:four_pt_fun_ksbasis_rot} we get
{{\begin{equation}
\begin{aligned}
\tilde{F}_{\substack{ R{\mathbf{k}_1a', R\mathbf{k}_2 b'} \\ {R\mathbf{k}_3c', R\mathbf{k}_4d'}}} &(\omega, \omega', \omega'')  =  \sum_{a, b, c, d}\Big\{ \ \tilde{F}_{\substack{ {\mathbf{k}_1a, \mathbf{k}_2 b} \\ {\mathbf{k}_3c, \mathbf{k}_4d}}} (\omega, \omega', \omega'') \\&
    \mathcal{D}^{\vphantom{*}}_{\mathbf{k}_1,a'a}(g) \mathcal{D}^*_{\mathbf{k}_3,c'c}(g)
    \mathcal{D}^*_{\mathbf{k}_2,b'b}(g) \mathcal{D}^{\vphantom{*}}_{\mathbf{k}_4,d'd}(g) \Big \}.
    \label{eq:F_symm}
\end{aligned}
\end{equation}
}}

Equation~\eqref{eq:F_symm} motivates us to define the matrices $\mathcal{U}(g)$ for all $g = \{R \mid \bm{\tau}\} \in \mathcal{G}$. They are given by  
\begin{equation}
    \mathcal{U}_{\substack{ {\mathbf{k}'_1a', \mathbf{k}'_2 b'} \\ {\mathbf{k}_1a\hspace{0.25em}, \mathbf{k}_2b\hspace{0.25em}}}}(g) =  \mathcal{D}^{\vphantom{*}}_{\mathbf{k}^{\vphantom{*}}_1,a'a}(g) \mathcal{D}^*_{\mathbf{k}^{\vphantom{*}}_2,b'b}(g) \delta^{\vphantom{*}}_{R\mathbf{k}^{\vphantom{*}}_1,\mathbf{k}_1'} \delta^{\vphantom{*}}_{R\mathbf{k}^{\vphantom{*}}_2,\mathbf{k}_2'}.
    \label{eq:exe_rep_mat}
\end{equation}
Similarly, in the case of time-reversal symmetry, the matrix $\mathcal{U}(\mathscr{T})$ is given by  
\begin{equation}
    \mathcal{U}_{\substack{ {\mathbf{k}'_1a', \mathbf{k}'_2 b'} \\ {\mathbf{k}_1a\hspace{0.25em},\mathbf{k}_2b\hspace{0.25em}}}}(\mathscr{T}) =  \mathcal{D}^{\vphantom{*}}_{\mathbf{k}^{\vphantom{*}}_1,a'a}(\mathscr{T}) \mathcal{D}^*_{\mathbf{k}^{\vphantom{*}}_2,b'b}(\mathscr{T}) \delta^{\vphantom{*}}_{-\mathbf{k}^{\vphantom{*}}_1,\mathbf{k}_1'} \delta^{\vphantom{*}}_{-\mathbf{k}^{\vphantom{*}}_2,\mathbf{k}_2'} \hat{\mathscr{K}},
    \label{eq:exe_rep_mat_trev}
\end{equation}
where $\hat{\mathscr{K}}$ is the complex conjugation operator introduced to ensure the anti-unitary property of time-reversal symmetry.

This allows us to conveniently express Eq.~\eqref{eq:F_symm} as 
\begin{equation}
    \mathcal{F} = \mathcal{U} \mathcal{F} \mathcal{U}^\dagger,
    \label{eq:UFUdagger}
\end{equation}
where $\mathcal{F}$ denotes the matrix $\tilde{F}_{\substack{ {\mathbf{k}_1a, \mathbf{k}_2 b} \\ {\mathbf{k}_3c, \mathbf{k}_4d}}} (\omega, \omega', \omega'')$. It is important to note that we rely on the closure property of the $\mathbf{k}$-point set under all group operations when writing $\mathcal{F} = \mathcal{U} \mathcal{F} \mathcal{U}^\dagger$. If this closure property is not satisfied (as can occur, for instance, with certain shifted $\mathbf{k}$-grids), the symmetry properties are lost, leading to artifacts such as the breaking of degeneracies. For example, in Refs.~\cite{PhysRevLett.96.026402,PhysRevLett.100.189701}, the use of a shifted grid that did not respect this property resulted in the breaking of degeneracies between the first set of bright and dark excitons in $h$BN.

An important point to note is that since the $\mathcal{D}$ matrices are unitary, the $\mathcal{U}$ matrices are also unitary for spatial symmetries (see Appendix~\ref{section:Uisunitary} for proof). A central result of this work is that the set of all matrices $\mathcal{U}(g)$ for $g \in \mathcal{G}$ constitutes a linear unitary representation of the space group $\mathcal{G}$~(see Appendix~\ref{section:U_is_rep_of_G} for proof).

An obvious consequence of Eq.~\eqref{eq:F_symm} arises when considering the effect of pure lattice translations. If $g$ corresponds to a translation, i.e., $\mathbf{r} \to \mathbf{r} + \bm{\tau}$, which belongs to the translational subgroup $\mathcal{T}$, then  
\begin{equation}
    \mathcal{D}_{\mathbf{k},mn}(g) = \delta_{m,n} e^{-i\mathbf{k} \cdot \bm{\tau}}.
    \label{eq:Dmat_trans}
\end{equation}  
Substituting Eq.~\eqref{eq:Dmat_trans} into Eq.~\eqref{eq:F_symm}, we obtain that the matrix elements $\tilde{F}_{\substack{ {\mathbf{k}_1a, \mathbf{k}_2 b} \\ {\mathbf{k}_3c, \mathbf{k}_4d}}} (\omega, \omega', \omega'')$ are nonzero only if the following crystal momentum conservation condition is satisfied: 
\begin{equation}
    \mathbf{k}_3 - \mathbf{k}_4 = \mathbf{k}_1 - \mathbf{k}_2 + \mathbf{G} = \mathbf{Q}.
    \label{eq:exe_mom_consversion}
\end{equation}

Eq.~\eqref{eq:exe_mom_consversion} implies that the matrix $\mathcal{F}$, along with the $\mathcal{U}$ matrices, are block diagonal in the basis of two-particle electron-hole states, with each block labeled by the crystal transfer momentum $\mathbf{Q}$. This crystal momentum corresponds to a one-dimensional representation of the subgroup $\mathcal{T}$ (analogous to the electronic Hamiltonian) and gives rise to the very well-known concept of dispersion (such as exciton dispersion)~\cite{PhysRevB.88.155113}. This implies that each block of $\mathcal{F}$  can be written as
\begin{equation}
\begin{aligned}
    \mathcal{F}^{(\mathbf{Q})}_{\vphantom{\substack{ {\mathbf{k}_1ab} \\ {\mathbf{k}_3cd}}}} = \tilde{F}^{(\mathbf{Q})}_{\substack{ {\mathbf{k}_1ab} \\ {\mathbf{k}_3cd}}}(\omega, \omega', \omega'') = \tilde{F}^{\vphantom{(\mathbf{Q})}}_{\substack{ {\mathbf{k}_1a, \mathbf{k}_1 - \mathbf{Q} b} \\ {\mathbf{k}_3c, \mathbf{k}_3 - \mathbf{Q}d}}} (\omega, \omega', \omega'').
\label{eq:exe_hami_block_diagonal}  
\end{aligned}
\end{equation}

Furthermore, Eq.~\eqref{eq:F_symm} shows that the block matrices $\mathcal{F}^{(\mathbf{Q})}$ and $\mathcal{F}^{(R\mathbf{Q})}$ are related by a similarity transformation,  
\begin{equation}
    \mathcal{F}^{(R\mathbf{Q})} = \mathcal{U}^{(\mathbf{Q})} \, \mathcal{F}^{(\mathbf{Q})} \, (\mathcal{U}^{(\mathbf{Q})})^\dagger,
    \label{eq:Q2rq}
\end{equation}
where the matrices $\mathcal{U}^{(\mathbf{Q})}(g)$ are defined as
\begin{equation}
\begin{aligned}
         \mathcal{U}^{(\mathbf{Q})}_{\vphantom{\substack{ {\mathbf{k}_1ab} \\ {\mathbf{k}_3cd}}}} &= \mathcal{U}^{(\mathbf{Q})}_{\substack{ {\mathbf{k}_1 ab} \\ {\mathbf{k}_3 cd}}}(g) 
         = \mathcal{U}^{\vphantom{(\mathbf{Q})}}_{\substack{ {\mathbf{k}_1a ,\mathbf{k}_1 -R\mathbf{Q}, b} \\ {\mathbf{k}_3c ,\mathbf{k}_3 -\hphantom{R}\mathbf{Q}, d}}}(g) \\
         &= \mathcal{D}^{\vphantom{*}}_{\mathbf{k}_3,ac}(g) \, \mathcal{D}^*_{\mathbf{k}_3-\mathbf{Q},bd}(g) \, \delta^{\vphantom{*}}_{R\mathbf{k}_3,\mathbf{k}_1} \ .
    \label{eq:exe_Umat_block_diago}
\end{aligned}
\end{equation}

For time-reversal symmetry $\mathscr{T}$,
\begin{equation}
\begin{aligned}
         \mathcal{U}^{(\mathbf{Q})}_{\vphantom{\substack{ {\mathbf{k}_1ab} \\ {\mathbf{k}_3cd}}}} &= \mathcal{U}^{(\mathbf{Q})}_{\substack{ {\mathbf{k}_1 ab} \\ {\mathbf{k}_3 cd}}}(\mathscr{T}) 
         = \mathcal{U}^{\vphantom{(\mathbf{Q})}}_{\substack{ {\mathbf{k}_1a ,\mathbf{k}_1 +\mathbf{Q}, b} \\ {\mathbf{k}_3c ,\mathbf{k}_3 -\mathbf{Q}, d}}}(\mathscr{T}) \\
         &= \mathcal{D}^{\vphantom{*}}_{\mathbf{k}_3,ac}(\mathscr{T}) \, \mathcal{D}^*_{\mathbf{k}_3-\mathbf{Q},bd}(\mathscr{T}) \, \delta^{\vphantom{*}}_{-\mathbf{k}_3,\mathbf{k}_1} \, \hat{\mathscr{K}} \ .
    \label{eq:exe_Umat_block_diago_trev}
\end{aligned}
\end{equation}

If $g$ belongs to the little group of $\mathbf{Q}$ i.e $R\mathbf{Q} = \mathbf{Q} + \mathbf{G}$, then we obtain
\begin{equation}
    \mathcal{F}^{(\mathbf{Q})} = \mathcal{U}^{(\mathbf{Q})} \mathcal{F}^{(\mathbf{Q})} (\mathcal{U}^{(\mathbf{Q})})^\dagger.
    \label{eq:Q2rq_lg}
\end{equation}

Now, we apply the formalism developed above to the BSE. The BSE is given by~\cite{PhysRevB.62.4927,PhysRevLett.81.2312,PhysRevB.21.4656,PhysRevLett.80.4510,sven_thesis}
\begin{equation}
    \begin{aligned}
    L(1,2;3,4) = L_0(1,2;3,4) + 
    &\int  d(5,6,7,8)  \ \Big \{ L_0(1,2;5,6) \\ & \times K(5,6;7,8) L(7,8;3,4) \Big \}
    \label{eq:bse}
    \end{aligned}
\end{equation}
where $L(1,2;3,4)$ is the electron-hole correlation function, and $L_0(1,2;5,6)$ is the independent-particle correlation function.

The electron-hole interaction kernel $K(5,6;7,8)$ is given by~\cite{Stefanucci2013}
\begin{equation}
    K(5,6;7,8) = \frac{ \delta \{\Sigma(5,6) + \delta(5,6)v_{H}(5) \} }{\delta G(7,8)},
    \label{eq:kernel_def}
\end{equation}
where $\Sigma(5,6)$ and $v_{H}(5)$ are the self-energy and the Hartree potential, respectively, and $G(7,8)$ is the single-particle Green's function. Within the GW approximation for the one-particle self-energy, along with the assumption that $\frac{\delta W(5,6)}{\delta G(7,8)} \approx 0$, the interaction kernel $K(5,6;7,8)$ can be expressed as the sum of the attractive screened Coulomb interaction and the repulsive bare exchange interaction, and is given by~\cite{PhysRevB.34.5390}:
\begin{equation}
\begin{aligned}
    K(5,6;7,8) = & \ iW(5,6) \delta(5,7) \delta(6,8) \\ &
    - iv(5,7) \delta(5,6) \delta(7,8),
    \label{eq:bse_k}
\end{aligned}
\end{equation}
where $v(5,7)$ and $W(5,6)$ are the bare and screened Coulomb potentials, respectively.

The standard procedure for solving the BSE involves substituting Eq.~\eqref{eq:bse_k} into Eq.~\eqref{eq:bse}, performing a Fourier transform of Eq.~\eqref{eq:bse} into the time domain, and expanding both $L(7,8;3,4)$ and $K(5,6;7,8)$ in a single-particle basis, as done in Eq.~\eqref{eq:four_pt_fun_ksbasis}~\cite{PhysRevLett.81.2312,PhysRevB.62.4927,PhysRevLett.80.4510}. 
Due to the invariance of the four-point functions $L(7,8;3,4)$ and $K(5,6;7,8)$ under space group operations (see Appendix~\ref{section:bse_kernel_invar} for proof), their Fourier transform in the single-particle basis follows Eq.~\eqref{eq:F_symm}. This implies that the effective two-particle BSE in the frequency domain can be solved separately for each momentum transfer $\mathbf{Q}$.  

Assuming that the electron-hole interaction kernel is static, i.e., without frequency dependence, the effective two-particle BSE for each momentum transfer $\mathbf{Q}$ is written as (see Refs.~\cite{RevModPhys.74.601,10.1063/1.3065669,sven_thesis} for derivation):
\begin{equation}
    (\mathcal{L}^\mathbf{Q})^{-1}(\omega) = (\mathcal{L}^\mathbf{Q}_0)^{-1}(\omega) - \mathcal{K}^\mathbf{Q},
    \label{eq:bse_simplified}
\end{equation}
where $\mathcal{L}^\mathbf{Q}$ and $\mathcal{K}^\mathbf{Q}$ are the matrices defined in Eq.~\eqref{eq:exe_hami_block_diagonal} for the two-particle correlation function and the electron-hole interaction kernel defined in Eq.~\eqref{eq:bse_k}, respectively. The independent particle correlation function $\mathcal{L}_0^\mathbf{Q}(\omega)$ is diagonal in the KS basis and can be approximated as~\cite{PhysRevB.62.4927,PhysRevLett.81.2312,PhysRevB.21.4656,PhysRevLett.80.4510,10.1063/1.3065669,sven_thesis}
\begin{equation}
\begin{aligned}
    \mathcal{L}_0^\mathbf{Q}(\omega)  = (\tilde{L}^{(\mathbf{Q})}_0)_{\substack{ {\mathbf{k}_1ab} \\ {\mathbf{k}_3cd}}}(\omega) \approx &\ i\delta_{a,c}\delta_{b,d} \delta_{\mathbf{k_1},\mathbf{k_3}} \\ & \times
    \frac{(f_{\mathbf{k_3},c} - f_{\mathbf{k_3}-\mathbf{Q},d})}{\omega - (\varepsilon_{\mathbf{k_3},c} - \varepsilon_{\mathbf{k_3}-\mathbf{Q},d} )},
    \label{eq:LQ0}
\end{aligned}
\end{equation}
where $f_{\mathbf{k_3},c}$ and $f_{\mathbf{k_3}-\mathbf{Q},d}$ are the occupation factors, and $\varepsilon_{\mathbf{k_3},c}$ and $\varepsilon_{\mathbf{k_3}-\mathbf{Q},d}$ are the single-particle energies for states $\mathbf{k_3},c$ and $\mathbf{k_3}-\mathbf{Q},d$, respectively.

Substituting Eq.~\eqref{eq:LQ0} into Eq.~\eqref{eq:bse_simplified}, we obtain
\begin{equation}
    \tilde{L}^{(\mathbf{Q})}_{\substack{ {\mathbf{k}_1ab} \\ {\mathbf{k}_3cd}}}(\omega) = \frac{i(f_{\mathbf{k_3},c} - f_{\mathbf{k_3}-\mathbf{Q},d})}{\omega - \tilde{H}^{(\mathbf{Q})}_{\substack{ {\mathbf{k}_1ab} \\ {\mathbf{k}_3cd}}}},
    \label{eq:bse_HQdef}
\end{equation}
where the two-particle exciton Hamiltonian $\mathcal{H}^{(\mathbf{Q})}$ is given by~\cite{PhysRevLett.81.2312}
\begin{equation}
\begin{aligned}
     \mathcal{H}^{(\mathbf{Q})}  = \tilde{H}^{(\mathbf{Q})}_{\substack{ {\mathbf{k}_1ab} \\ {\mathbf{k}_3cd}}} = & \ \delta_{a,c}\delta_{b,d} \delta_{\mathbf{k_1},\mathbf{k_3}}(\varepsilon_{\mathbf{k_3},c} - \varepsilon_{\mathbf{k_3}-\mathbf{Q},d} )  \\& + i(f_{\mathbf{k_3},c} - f_{\mathbf{k_3}-\mathbf{Q},d}) \tilde{K}^{(\mathbf{Q})}_{\substack{ {\mathbf{k}_1ab} \\ {\mathbf{k}_3cd}}}.
    \label{eq:bs_ham}
\end{aligned}
\end{equation}
The matrix $\mathcal{H}^{(\mathbf{Q})}$ defined in Eq.~\eqref{eq:bs_ham} is, in general, non-Hermitian ~\cite{Martin_Reining_Ceperley_2016}. Its eigenvalues and left/right eigenvectors correspond to the energies and left/right eigenstates of excitonic states in the electron-hole basis with momentum transfer $\mathbf{Q}$.

It is important to note that, although $\mathcal{H}^{(\mathbf{Q})}$ in Eq.~\eqref{eq:bs_ham} contains the factor $(f_{\mathbf{k_3},c} - f_{\mathbf{k_3}-\mathbf{Q},d})$, it nevertheless satisfies Eq.~\eqref{eq:Q2rq} owing to the block-diagonal structure of the $\mathcal{D}$ matrices within the degenerate subspace. As a result, the excitonic Hamiltonians at $\mathbf{Q}$ and $R\mathbf{Q}$ ($-\mathbf{Q}$ in the case of time-reversal symmetry) are connected through

\begin{equation}
    \mathcal{H}^{(R\mathbf{Q})} = \mathcal{U}^{(\mathbf{Q})} \mathcal{H}^{(\mathbf{Q})} (\mathcal{U}^{(\mathbf{Q})})^\dagger.
    \label{eq:Q2rq_H}
\end{equation}

Equation~\eqref{eq:Q2rq_H} ensures that $\mathcal{H}^{(R\mathbf{Q})}$ and $\mathcal{H}^{(\mathbf{Q})}$ share same eigenvalues, with their eigenvectors connected via
\begin{equation}
\begin{aligned}
    &\mathcal{A}^{(R\mathbf{Q})} \equiv \mathcal{U}^{(\mathbf{Q})} \mathcal{A}^{(\mathbf{Q})},\\
    &\tilde{\mathcal{A}}^{(R\mathbf{Q})} \equiv \mathcal{U}^{(\mathbf{Q})}\tilde{\mathcal{A}}^{(\mathbf{Q})},
    \label{eq:Q2rq_H_rightev}
\end{aligned}
\end{equation}
where the columns of $\mathcal{A}^{(R\mathbf{Q})}$ / $\tilde{\mathcal{A}}^{(R\mathbf{Q})}$ and $\mathcal{A}^{(\mathbf{Q})}$ / $\tilde{\mathcal{A}}^{(\mathbf{Q})}$ denote the right/left eigenvectors of $\mathcal{H}^{(R\mathbf{Q})}$ and $\mathcal{H}^{(\mathbf{Q})}$, respectively. We use $\equiv$ to indicate that $\mathcal{A}^{(R\mathbf{Q})}$ and $\mathcal{U}^{(\mathbf{Q})} \mathcal{A}^{(\mathbf{Q})}$ are equivalent up to a phase or a unitary rotation.

It is worth emphasizing that Eq.~\eqref{eq:Q2rq_H_rightev} shows how a symmetry operation transforms an exciton wavefunction with momentum transfer $\mathbf{Q}$ into one with momentum transfer $R\mathbf{Q}$, while preserving its eigenvalue. This implies that for a right excitonic wavefunction with momentum transfer \(\mathbf{Q}\), which is expressed as~\cite{PhysRevLett.81.2312,PhysRevB.62.4927,PhysRevLett.80.4510}  
\begin{equation}
\Psi_S^{\mathbf{Q}}(\mathbf{r}_e,\mathbf{r}_h) = \sum_{\mathbf{k}ij} \mathcal{A}^{S,(\mathbf{Q})}_{\mathbf{k}ij} \ \phi^{\vphantom{-\mathbf{Q}}}_{\mathbf{k}i}(\mathbf{r}_e) \ \phi^*_{\mathbf{k}-\mathbf{Q},j}(\mathbf{r}_h),
\label{eq:exe_stateInR111}    
\end{equation}
the action of a symmetry operator on \(\Psi_S^{\mathbf{Q}}(\mathbf{r}_e,\mathbf{r}_h)\) can be written as
\begin{equation}
\begin{aligned}
\{R \mid \bm{\tau}\}\Psi_S^{\mathbf{Q}}(\mathbf{r}_e,\mathbf{r}_h) = \sum_{\mathbf{k}iji'j'\mathbf{k}'} & \Big \{ \mathcal{U}^{(\mathbf{Q})}_{
\substack{ {\mathbf{k}'i'j'} \\ {\mathbf{k}\hphantom{'}i\hphantom{'}j\hphantom{'}}}}(g) \mathcal{A}^{S,(\mathbf{Q})}_{\mathbf{k}ij} \\ & \times \phi^{\vphantom{*}}_{\mathbf{k}'i'}(\mathbf{r}_e) \ \phi^*_{\mathbf{k}'-R\mathbf{Q},j'}(\mathbf{r}_h) \Big\}.
\end{aligned}
\label{eq:exe_stateInR222}    
\end{equation}
Similarly, the action of the time-reversal operator can be expressed as
\begin{equation}
\begin{aligned}
\mathscr{T}\Psi_S^{\mathbf{Q}}(\mathbf{r}_e,\mathbf{r}_h) = \sum_{\mathbf{k}iji'j'\mathbf{k}'} & \Big \{ \Big ( \mathcal{U}^{(\mathbf{Q})}_{
\substack{ {\mathbf{k}'i'j'} \\ {\mathbf{k}\hphantom{'}i\hphantom{'}j\hphantom{'}}}}(\mathscr{T}) \mathcal{A}^{S,(\mathbf{Q})}_{\mathbf{k}ij} \Big) \\ & \times \phi^{\vphantom{*}}_{\mathbf{k}'i'}(\mathbf{r}_e) \ \phi^*_{\mathbf{k}'+\mathbf{Q},j'}(\mathbf{r}_h) \Big\},
\end{aligned}
\label{eq:exe_stateInR222_1}
\end{equation}
where \(\mathbf{r}_e\) and \(\mathbf{r}_h\) denote the positions of  the electron and the hole, respectively.
In Eq.~\eqref{eq:exe_stateInR222_1}, parentheses are included to indicate that the complex conjugation operator $\mathscr{K}$ in $\mathcal{U}^{(\mathbf{Q})}_{
\substack{ {\mathbf{k}'i'j'} \\ {\mathbf{k}\hphantom{'}i\hphantom{'}j\hphantom{'}}}}(\mathscr{T})$, as defined in Eq.~\eqref{eq:exe_rep_mat_trev}, acts solely on the $\mathcal{A}$ vectors.

Since $\{R \mid \bm{\tau}\}\Psi_S^{\mathbf{Q}}(\mathbf{r}_e,\mathbf{r}_h)$ and $\Psi_S^{R\mathbf{Q}}(\mathbf{r}_e,\mathbf{r}_h)$ correspond to the same states (for time-reversal $\mathscr{T}\Psi_S^{\mathbf{Q}}(\mathbf{r}_e,\mathbf{r}_h)$ and $\Psi_S^{-\mathbf{Q}}(\mathbf{r}_e,\mathbf{r}_h)$), they must differ by a phase (or by a rotation matrix for degenerate excitons), which is given by
\begin{equation}
\begin{aligned}
    \{R \mid \bm{\tau}\}\Psi_S^{\mathbf{Q}}(\mathbf{r}_e,\mathbf{r}_h) = \sum_{S'}\mathscr{D}^{\vphantom{-\mathbf{Q}}}_{\mathbf{Q},S'S}(g) \Psi_{S'}^{R\mathbf{Q}}(\mathbf{r}_e,\mathbf{r}_h),
    \label{eq:Exe_dmat_symm1}
\end{aligned}
\end{equation}
\begin{equation}
\begin{aligned}
    \mathscr{T}\Psi_S^{\mathbf{Q}}(\mathbf{r}_e,\mathbf{r}_h) = \sum_{S'} \mathscr{D}^{\vphantom{-\mathbf{Q}}}_{\mathbf{Q},S'S}(\mathscr{T}) \Psi_{S'}^{-\mathbf{Q}}(\mathbf{r}_e,\mathbf{r}_h),
    \label{eq:Exe_dmat_symm2}
\end{aligned}
\end{equation}
where $\mathscr{D}_{\mathbf{Q},S'S}(g)/\mathscr{D}_{\mathbf{Q},S'S}(\mathscr{T})$ is a rotation matrix which is block diagonal in degenerate space i.e, $\mathscr{D}_{\mathbf{Q}}(g) = \bigoplus_i \mathscr{D}^i_{\mathbf{Q}}(g)$. If we choose the left eigenvectors such that the overlap matrix between the left and right eigenvectors is an identity matrix~\cite{Ashida2020Jul} i.e., $(\mathcal{\tilde{A}}^{(\mathbf{Q})})^\dagger \mathcal{A}^{(\mathbf{Q})} = I$, then $\mathscr{D}_{\mathbf{Q},S'S}(g)$ and $\mathscr{D}_{\mathbf{Q},S'S}(\mathscr{T})$  are given by
\begin{equation}
\begin{aligned}
\mathscr{D}_{\mathbf{Q},S'S}(g) = \sum_{\mathbf{k}iji'j'\mathbf{k}'} \mathscr{U}^{(\mathbf{Q})}_{\substack{ {\mathbf{k}'i'j'} \\ {\mathbf{k}\hphantom{'}i\hphantom{'}j\hphantom{'}}}}(g) \mathcal{A}^{S,(\mathbf{Q})}_{\mathbf{k}ij} \left(\mathcal{\tilde{A}}^{,S'(R\mathbf{Q})}_{\mathbf{k}'i'j'}\right)^* \\
= \sum_{\mathbf{k}iji'j'} \mathcal{D}^{\vphantom{*}}_{\mathbf{k},i'i}(g) (\mathcal{D}^{\vphantom{*}}_{\mathbf{k}-\mathbf{Q},j'j}(g))^* \mathcal{A}^{S,(\mathbf{Q})}_{\mathbf{k}ij} \left(\mathcal{\tilde{A}}^{,S'(R\mathbf{Q})}_{R\mathbf{k}i'j'}\right)^* ,\\
\end{aligned}
\label{eq:exe_dmat_nonher}
\end{equation}
\begin{equation}
\begin{aligned}
\mathscr{D}_{\mathbf{Q},S'S}(\mathscr{T}) = \sum_{\mathbf{k}iji'j'\mathbf{k}'} \Big ( \mathscr{U}^{(\mathbf{Q})}_{\substack{ {\mathbf{k}'i'j'} \\ {\mathbf{k}\hphantom{'}i\hphantom{'}j\hphantom{'}}}}(\mathscr{T}) \mathcal{A}^{S,(\mathbf{Q})}_{\mathbf{k}ij} \Big )\left(\mathcal{\tilde{A}}^{,S'(R\mathbf{Q})}_{\mathbf{k}'i'j'}\right)^* \\
= \sum_{\mathbf{k}iji'j'} \mathcal{D}^{\vphantom{*}}_{\mathbf{k},i'i}(\mathscr{T}) \left(\mathcal{D}^{\vphantom{*}}_{\mathbf{k}-\mathbf{Q},j'j}(\mathscr{T})\right)^* \left(\mathcal{A}^{S,(\mathbf{Q})}_{\mathbf{k}ij}\right)^*\left(\mathcal{\tilde{A}}^{,S'(-\mathbf{Q})}_{-\mathbf{k}i'j'}\right)^* .
\end{aligned}
\label{eq:exe_dmat_nonher2}
\end{equation}
Equations \eqref{eq:exe_dmat_nonher} and \eqref{eq:exe_dmat_nonher2} constitute another central result of this work. The $\mathscr{D}_{\mathbf{Q}}(g)/\mathscr{D}_{\mathbf{Q}}(\mathscr{T})$ matrices can be obtained through straightforward post-processing of standard \emph{ab initio} calculations, as conventional BSE codes output the eigenvectors $\mathcal{A}$, and $\mathcal{D}_{\mathbf{k}}(g)$ can be computed from the Kohn–Sham wavefunctions using Eqs.~\eqref{eq:dmats} and~\eqref{eq:dmats_trev}. 

If $g$ is in the little group of $\mathbf{Q}$, then $\mathscr{D}_{\mathbf{Q},S'S}(g)$ corresponds to representation matrices of symmetries in the excitonic basis. It should be noted that, unlike the $\mathcal{D}$ matrices for electrons (when spin is not neglected), the $\mathscr{D}$ matrices for excitons respect the space group multiplication, as the $\mathcal{U}^{(\mathbf{Q})}$ form a linear representation of the space group.

In order to obtain the irreducible labels for the excitonic states, we decompose the little group $\mathcal{G}_{\mathbf{Q}}$ of $\mathbf{Q}$ into left cosets of the subgroup $\mathcal{T}$ in $\mathcal{G}_{\mathbf{Q}}$, i.e.,
\begin{equation}
    \mathcal{G}_{\mathbf{Q}} \equiv \bigcup_{i=1}^{n_{\mathbf{Q}}} g^{\mathbf{Q}}_{i}\mathcal{T},
    \label{eq:Gcosetdecom11111}
\end{equation}
where $g^{\mathbf{Q}}_{i}$ are the coset representatives, $g^{\mathbf{Q}}_{i}\mathcal{T}$ represents the left cosets of $\mathcal{T}$ in $\mathcal{G}_{\mathbf{Q}}$, and $n_{\mathbf{Q}}$ is the index of $\mathcal{T}$ in $\mathcal{G}_{\mathbf{Q}}$.  The quotient group $\mathcal{G}_{\mathbf{Q}}/\mathcal{T}$ is isomorphic to the little point group $\mathcal{P}_{\mathbf{Q}}$, which is a subgroup of the point group $\mathcal{P}$ of the crystal~\cite{el2008symmetry,dresselhaus2007group}.

The set of coset representatives $\{g^{\mathbf{Q}}_{i}\}$ is obtained by removing the pure lattice translation (excluding fractional translations) from the little group symmetries $\mathcal{G}_\mathbf{Q}$. This set does not necessarily form a group,  which allows for a slight redefinition of the representation matrices. The new, redefined exciton representation matrix for an element $g$ is given by  
\begin{equation}
    \bar{\mathscr{D}}_{\mathbf{Q}}(g) = \mathscr{D}_{\mathbf{Q}}(g) e^{i\mathbf{Q}\cdot \boldsymbol{\tau}}.
    \label{eq:Dmat_exe_newrep11}
\end{equation}

The new representation matrices are identical for all elements within a given coset. The group multiplication rule for two elements $g_1 = \{ R_1 \ | \ \boldsymbol{\tau}_1 \}$ and $g_2 = \{ R_2 \ | \ \boldsymbol{\tau}_2 \}$ from two different cosets is given by  
\begin{equation}
    \bar{\mathscr{D}}_{\mathbf{Q}}(g_1) \bar{\mathscr{D}}_{\mathbf{Q}}(g_2) = \bar{\mathscr{D}}_{\mathbf{Q}}(g_1\cdot g_2) e^{-i\mathbf{G}_0\cdot \boldsymbol{\tau}_2},
    \label{eq:new_rep_group_mul}
\end{equation}
where $\mathbf{G}_0 = R_1^{-1}\mathbf{Q} - \mathbf{Q}$. If $\mathbf{G}_0 = 0$, or if there are no fractional translations in the little group $\mathcal{G}_{\mathbf{Q}}$ (more precisely if $e^{-i\mathbf{G}_0\cdot \boldsymbol{\tau}_2} = 1$), then $\bar{\mathscr{D}}_{\mathbf{Q}}$ corresponds to a linear representation; otherwise, it corresponds to a projective representation of the point group $\mathcal{P}_{\mathbf{Q}}$. In other words, $\mathscr{D}_{\mathbf{Q}}$ corresponds to a trivial projective representation of the point group if $\mathbf{G}_0 = 0$, in which case the representations can be mapped to ordinary representations via the phase transformation given in Eq.~\eqref{eq:new_rep_group_mul}, and to a non-trivial projective representation otherwise. This implies that, as in the case of phonons~\cite{RevModPhys.40.1}, the excitonic states can be labeled by the projective irreducible representations of the little point groups. For demonstration purposes, we consider only points where $e^{-i\mathbf{G}_0\cdot \boldsymbol{\tau}_2} = 1$, so that the trivial projective representations can be mapped to the ordinary representations of the point groups. Non-trivial projective representations can be worked out following the procedure described in Refs.~\cite{el2008symmetry,Kim1999Jun}.

To obtain the irreducible labels for the excitonic states, we consider each degenerate subspace block $\bar{\mathscr{D}}^i_{\mathbf{Q}}(g) = \mathscr{D}^i_{\mathbf{Q}}(g) e^{i\mathbf{Q}\cdot \boldsymbol{\tau}}$ of $\bar{\mathscr{D}}_{\mathbf{Q}}$ and decompose into the direct sum of irreducible representations i.e.,
\begin{equation}
    \bar{\mathscr{D}}^i_{\mathbf{Q}} = \bigoplus c_l \bar{\mathscr{D}}_{\mathbf{Q}}^{i;l},
    \label{eq:irrep_decom_exe1190}
\end{equation}
where $\bar{\mathscr{D}}_{\mathbf{Q}}^{i;l}$ corresponds to the $l^{\text{th}}$ irreducible representation of $\mathcal{P}_{\mathbf{Q}}$, and $c_l$ is its multiplicity. We use the standard orthogonality relation~\cite{el2008symmetry,dresselhaus2007group} for the characters to obtain $c_l$, which is given by
\begin{equation}
    c_l = \frac{1}{|G|} \sum_{g \in G} \chi^{(l)}(g)^* \chi(g),
\end{equation}
where $|G|$ is the order of the group, $\chi^{(l)}(g)$ is the characters of the $l^{\text{th}}$ irreducible representation, and $\chi(g)$ is the trace of $\bar{\mathscr{D}}^i_{\mathbf{Q}}$. 

\subsection{Simplification for Semiconductors}
Upto now, we have derived expressions without making any assumptions regarding the electronic structure of the crystal. However, in general, BSE is often used to study electron-hole excitations in semiconductors. This allows us to further simplify the expressions derived above. In the case of cold semiconductors, the exciton Hamiltonian defined in Eq.~\eqref{eq:bs_ham} can be expressed in a $2 \times 2$ block form, written as~\cite{Martin_Reining_Ceperley_2016}:
\begin{equation}
    \mathcal{H}_{2p}^{(\mathbf{Q})} = \begin{pmatrix}
    R^{(\mathbf{Q})} & C^{(\mathbf{Q})} \\
    -(C^{(\mathbf{Q})})^\dagger & D^{(\mathbf{Q})} 
    \end{pmatrix}
\label{eq:BSEMatrixbloc111}
\end{equation}
where $R^{(\mathbf{Q})}$ and $D^{(\mathbf{Q})}$ are known as the resonant and anti-resonant blocks, respectively, while $C^{(\mathbf{Q})}$ is the coupling block. The band transitions $(c,d) \rightarrow (a,b)$ for the $R^{(\mathbf{Q})}$ block are $(\tilde{c},\tilde{v}) \rightarrow (\tilde{c}',\tilde{v}')$, for the $D^{(\mathbf{Q})}$ block the transitions are $(\tilde{v},\tilde{c}) \rightarrow (\tilde{v}',\tilde{c}')$, and for the $C^{(\mathbf{Q})}$ block the transitions are $(\tilde{c},\tilde{v}) \rightarrow (\tilde{v}',\tilde{c}')$. Here, $\tilde{c}, \tilde{c}'$ are the indices of the conduction bands with occupation factor $f_{\mathbf{k},\tilde{c}/\tilde{c}'}=0$, and $\tilde{v}, \tilde{v}'$ are the indices of the valence bands with occupation factor $f_{\mathbf{k},\tilde{v}/\tilde{v}'}=1$. Often the coupling block $C$ is set to $0$, which is commonly referred to as the Tamm-Dancoff approximation~\cite{tda_ref}. Within the Tamm-Dancoff approximation (TDA), the Hamiltonian becomes Hermitian, and the eigenvectors form an orthogonal set.

By diagonalizing the matrix given in Eq.~\eqref{eq:BSEMatrixbloc111}, we obtain the excitonic energies and eigenstates:
\begin{equation}
   \begin{pmatrix}
    \hphantom{-}R^{(\mathbf{Q})}\hphantom{\dagger} & C^{(\mathbf{Q})} \\
    -\left(C^{(\mathbf{Q})}\right)^\dagger & D^{(\mathbf{Q})} 
    \end{pmatrix}
    \begin{pmatrix}
    X^{S;(\mathbf{Q})}  \\
    \left.Y^{S;(\mathbf{Q})}\right.^{\vphantom{\dagger}}
    \end{pmatrix}
    = \varepsilon^{S;(\mathbf{Q})} 
    \begin{pmatrix}
    X^{S;(\mathbf{Q})}  \\
    \left.Y^{S;(\mathbf{Q})}\right.^{\vphantom{\dagger}}
    \end{pmatrix},
\label{eq:BSEMatrixbloc_diagonalize}
\end{equation}
where 
$
 \begin{pmatrix}
X^{S;(\mathbf{Q})} \\
Y^{S;(\mathbf{Q})}
\end{pmatrix}
$
is the right excitonic wavefunction in the electron-hole basis with energy $\varepsilon^{S;(\mathbf{Q})}$.

Using Eq.~\eqref{eq:Q2rq_H}, we can then write the similarity transformation for the $\mathcal{H}_{2p}^{(\mathbf{Q})}$ as
\begin{equation}
    \mathcal{H}_{2p}^{(R\mathbf{Q})} = \mathscr{U}_{\vphantom{2p}}^{(\mathbf{Q})}(g) \mathcal{H}_{2p}^{(\mathbf{Q})} 
    \left(\mathscr{U}_{\vphantom{2p}}^{(\mathbf{Q})}(g)\right)^\dagger,
    \label{eq:H_mat_Q2rq}
\end{equation}
where $\mathscr{U}^{(\mathbf{Q})}(g)$ is defined as 
\begin{equation}
    \mathscr{U}^{(\mathbf{Q})}(g)  =  \begin{pmatrix}
    \mathcal{U}_1^{(\mathbf{Q})}(g)  & 0 \\
    0 & \mathcal{U}_2^{(\mathbf{Q})}(g) 
    \end{pmatrix},
    \label{eq:exe_Umat_def}
\end{equation}
with $\mathcal{U}_1^{(\mathbf{Q})}(g)$ and $\mathcal{U}_2^{(\mathbf{Q})}(g)$ given in Eq.~\eqref{eq:exe_Umat_block_diago}. Here we used $\mathcal{D}_{\mathbf{k},cv}(g) = 0$ as the valence and conduction bands are distinctly separated in energy from each other.
The band transitions $(c,d) \rightarrow (a,b)$ for the $\mathcal{U}_1^{(\mathbf{Q})}(g) $ block correspond to $(\tilde{c},\tilde{v}) \rightarrow (\tilde{c}',\tilde{v}')$, while for $\mathcal{U}_2^{(\mathbf{Q})}(g)$, they correspond to $(\tilde{v},\tilde{c}) \rightarrow (\tilde{v}',\tilde{c}')$. 

Similarly, for time-reversal symmetry $\mathscr{T}$, Eq.~\eqref{eq:H_mat_Q2rq} takes the following form:  
\begin{equation}
    \mathcal{H}_{2p}^{(-\mathbf{Q})} = \mathscr{U}_{\vphantom{2p}}^{(\mathbf{Q})}(\mathscr{T}) \mathcal{H}_{2p}^{(\mathbf{Q})} 
    \left(\mathscr{U}_{\vphantom{2p}}^{(\mathbf{Q})}(\mathscr{T})\right)^\dagger,
    \label{eq:H_mat_Q2rq_trev}
\end{equation}
where the matrix $\mathscr{U}^{(\mathbf{Q})}(\mathscr{T})$ is given by  
\begin{equation}
    \mathscr{U}^{(\mathbf{Q})}(\mathscr{T})  =  \begin{pmatrix}
    \mathcal{U}_1^{(\mathbf{Q})}(\mathscr{T})  & 0 \\
    0 & \mathcal{U}_2^{(\mathbf{Q})}(\mathscr{T}) 
    \end{pmatrix}.
    \label{eq:exe_Umat_def_trev}
\end{equation}

The action of $\mathscr{U}^{(\mathbf{Q})}(g)$ and $\mathscr{U}^{(\mathbf{Q})}(\mathscr{T})$ on the excitonic wavefunction is written as 
\begin{equation}
\begin{aligned}
\begin{pmatrix}
X'^{S';(R\mathbf{Q})}_{\vphantom{1}} \\
Y'^{S';(R\mathbf{Q})}_{\vphantom{2}}
\end{pmatrix} =
\mathscr{U}^{(\mathbf{Q})}(g)\begin{pmatrix}
X^{S;(\mathbf{Q})}_{\vphantom{1}} \\
Y^{S;(\mathbf{Q})}_{\vphantom{2}}
\end{pmatrix}
=
\begin{pmatrix}
\mathcal{U}_1^{(\mathbf{Q})}(g)X^{S;(\mathbf{Q})}_{\vphantom{1}} \\
\mathcal{U}_2^{(\mathbf{Q})}(g)Y^{S;(\mathbf{Q})}_{\vphantom{2}}
\end{pmatrix}
\end{aligned}
\label{eq:exe_wf_rotate_0}
\end{equation}
where 
\begin{eqnarray}
    X'^{S';(R\mathbf{Q})}_{R\mathbf{k}cv} & = & \sum_{c'v'} \mathcal{D}_{\mathbf{k},cc'}(g) (\mathcal{D}_{\mathbf{k}-\mathbf{Q},vv'}(g))^* X^{S;(\mathbf{Q})}_{\mathbf{k}c'v'},\nonumber \\
    Y'^{S';(R\mathbf{Q})}_{R\mathbf{k}vc} & = & \sum_{c'v'} (\mathcal{D}_{\mathbf{k}-\mathbf{Q},cc'})^*(g) \mathcal{D}_{\mathbf{k},vv'}(g) Y^{S;(\mathbf{Q})}_{\mathbf{k}v'c'}.
    \label{eq:exe_wf_rotate_1}
\end{eqnarray}
The $\mathscr{D}_{\mathbf{Q},S'S}(g)$ defined in Eq.~\eqref{eq:exe_dmat_nonher} are obtained by taking the inner product of the rotated wavefunctions $\begin{pmatrix}X'^{S';(R\mathbf{Q})} \\Y'^{S';(R\mathbf{Q})}\end{pmatrix}$ with the left eigenvector of the $R\mathbf{Q}$ eigenstates.

Within the TDA, positive eigenvalues have $Y = 0$ and negative eigenvalues have $X = 0$. This implies that, from Eq.~\eqref{eq:exe_wf_rotate_1}, the $\mathscr{D}_{\mathbf{Q},S'S}(g)$ matrices for positive eigenvalues are given by
\begin{equation}
\begin{aligned}
    \mathscr{D}_{\mathbf{Q},S'S}(g) =
    \sum_{c'v',c,v,\mathbf{k}} 
    \Big\{& \mathcal{D}_{\mathbf{k},cc'}(g) 
    \big(\mathcal{D}_{\mathbf{k}-\mathbf{Q},vv'}(g)\big)^* \\&
    X^{S;(\mathbf{Q})}_{\mathbf{k}c'v'} 
    \big(X^{S';(R\mathbf{Q})}_{R\mathbf{k}cv}\big)^*\Big\}.
\end{aligned}
    \label{eq:exe_wf_rotate_tda}
\end{equation}

\section{Exploiting symmetries for Computational Aspects}
One of the important aspects of symmetries in computational physics is that they can significantly reduce the computational resources required to calculate material properties.
In this section, we discuss how the symmetry properties described in this work can be used to improve the computational efficiency of solving the BSE.

\subsection{Expansion of exciton wavefunctions from the irreducible Brillouin zone to the full Brillouin zone}
With the recent development of workflows for evaluating exciton–phonon matrix elements which are required for computing excitonic lifetimes and optical spectra, it has become increasingly important to obtain excitonic wavefunctions throughout the full Brillouin zone~\cite{PhysRevB.99.081109,PhysRevLett.125.107401,PhysRevMaterials.7.024006,Chan2023May,Marini2024May,PhysRevLett.122.187401,Zanfrognini2023Nov}. A straightforward approach is to construct and diagonalize the Bethe–Salpeter Hamiltonian at every $\mathbf{Q}$ point. However, this procedure is extremely computationally demanding, since it requires handling very large Hamiltonians across the entire Brillouin zone. This cost can be reduced by solving the BSE only in the irreducible Brillouin zone and obtaining the remaining results using symmetry operations.
As shown in Eq.~\eqref{eq:Q2rq_H_rightev} and Eq.~\eqref{eq:Exe_dmat_symm1}, we obtain a rotated exciton eigenvector upon the action of a crystal symmetry on the excitonic wavefunction. If $R\mathbf{Q} \neq \mathbf{Q} + \mathbf{G}$, then we can choose the rotated exciton wavefunction by setting the phase matrix $\mathscr{D}_{\mathbf{Q},S'S}(g) = \delta_{S'S}$. This implies that the right and left exciton wavefunctions for the wave vector $R\mathbf{Q}$ (or $-\mathbf{Q}$ in the case of time-reversal) are given by
\begin{eqnarray}
\mathcal{A}^{S,(R\mathbf{Q})}_{R\mathbf{k}i'j'}
& = & \sum_{\mathbf{k}ij} \mathcal{D}_{\mathbf{k},i'i}(g) (\mathcal{D}_{\mathbf{k}-\mathbf{Q},j'j}(g))^* \mathcal{A}^{S,(\mathbf{Q})}_{\mathbf{k}ij} ,\label{eq:exe_wf_ibz_BZ_trev} \\
\mathcal{\tilde{A}}^{S,(R\mathbf{Q})}_{R\mathbf{k}i'j'}
& = & \sum_{ij} \mathcal{D}_{\mathbf{k},i'i}(g) (\mathcal{D}_{\mathbf{k}-\mathbf{Q},j'j}(g))^* \mathcal{\tilde{A}}^{S,(\mathbf{Q})}_{\mathbf{k}ij} , \nonumber \\
\mathcal{A}^{S,(-\mathbf{Q})}_{-\mathbf{k}i'j'}
& = & \sum_{\mathbf{k}ij} \mathcal{D}_{\mathbf{k},i'i}(\mathscr{T}) (\mathcal{D}_{\mathbf{k}-\mathbf{Q},j'j}(\mathscr{T}))^* \left(\mathcal{A}^{S,(\mathbf{Q})}_{\mathbf{k}ij}\right)^* ,\nonumber \\
\mathcal{\tilde{A}}^{S,(-\mathbf{Q})}_{-\mathbf{k}i'j'}
& = & \sum_{ij} \mathcal{D}_{\mathbf{k},i'i}(\mathscr{T}) \left(\mathcal{D}_{\mathbf{k}-\mathbf{Q},j'j}(\mathscr{T})\right)^* \left(\mathcal{\tilde{A}}^{S,(\mathbf{Q})}_{\mathbf{k}ij}\right)^* .
\nonumber
\end{eqnarray}
For the case of semiconductors, we can set $X'^{S;(R\mathbf{Q})} = X^{S;(R\mathbf{Q})}$ and $Y'^{S;(R\mathbf{Q})} = Y^{S;(R\mathbf{Q})}$ in Eq.~\eqref{eq:exe_wf_rotate_1}. This allows us to obtain the exciton wavefunctions in the full Brillouin zone by \emph{only} computing the wavefunctions in the irreducible Brillouin zone.

It is important to note that multiple symmetry operations can map $\mathbf{Q}$ to $R\mathbf{Q}$ or $-\mathbf{Q}$. This implies that by fixing $\mathscr{D}_{\mathbf{Q},S'S}(g) = \delta_{S'S}$, we are choosing a particular phase convention~(or gauge). It is crucial to maintain gauge consistency when computing physical observables involving the exciton wavefunctions in the full Brillouin zone.
\subsection{Exploiting symmetries in the construction of the BSE Hamiltonian}
One of the most demanding steps in performing BSE calculations is the construction of the large BSE Hamiltonian matrix. Here, we show how symmetries can be used to reduce the computational cost of constructing the BSE Hamiltonian.
Consider the BSE Hamiltonian $\mathcal{H}^{(\mathbf{Q})}$ and a symmetry element 
$g = \{ R \mid \boldsymbol{\tau} \} \in \mathcal{G}_{\mathbf{Q}}$, i.e., 
$R\mathbf{Q} \equiv \mathbf{Q}$. Then, from Eq.~\eqref{eq:Q2rq_H} and 
Eq.~\eqref{eq:exe_Umat_block_diago}, we obtain
{{\begin{equation}
\begin{aligned}
\mathcal{H}^{(\mathbf{Q})}_{\substack{ R{\mathbf{k}'a' b'} \\ {R\mathbf{k}\hphantom{'}c'd'}}}  =  \sum_{a, b, c, d} 
\Big\{ &
 \mathcal{H}^{(\mathbf{Q})}_{\substack{ {\mathbf{k}'a b} \\ {\mathbf{k}\hphantom{'}cd}}}  \ 
    \mathcal{D}_{\mathbf{k}',a'a}(g) \ 
    \mathcal{D}^*_{\mathbf{k},c'c}(g) \\&
    \mathcal{D}^*_{\mathbf{k}'-\mathbf{Q},b'b}(g) \ \mathcal{D}_{\mathbf{k}-\mathbf{Q},d'd}(g) \Big \}.
    \label{eq:H_symm_rotation}
\end{aligned}
\end{equation}
}}
With Eq.~\eqref{eq:H_symm_rotation}, one can reduce the number of matrix elements that need to be computed by explicitly evaluating the bare and screened Coulomb matrix elements, while obtaining the rest from symmetry relations. This can drastically reduce the time required to construct the entire BSE Hamiltonian, particularly at the $\Gamma$ point where the full point group symmetry, including time-reversal (if present), can be exploited.

\subsection{Projection operators and block diagonalization}
After constructing the BSE Hamiltonian, one needs to diagonalize the matrix to obtain the eigenvalues and eigenvectors. Since these matrices are often very large, diagonalization can become the main bottleneck in BSE calculations. This cost can be reduced by transforming the BSE Hamiltonian into a block-diagonal form using a symmetry-adapted basis. To construct such a basis, one must first define the appropriate projection operators~\cite{el2008symmetry,Kim1999Jun,tung1985group}.

Consider the little point group $\mathcal{P}_{\mathbf{Q}}$, which is isomorphic to the quotient group $\mathcal{G}_{\mathbf{Q}}/\mathcal{T}$. Using the definition of the representation matrices $\mathcal{U}^{(\mathbf{Q})}(g)$ in Eq.~\eqref{eq:exe_Umat_block_diago}, we define projection operators $\hat{P}^{(l)}$ for the $l^{\text{th}}$ projective irreducible representation as~\cite{Altmann1989Mar}
\begin{equation}
    \hat{P}^{(l)}_{ij} = \frac{l_n}{|\mathcal{P}_{\mathbf{Q}}|} \sum_{g \in \mathcal{P}_{\mathbf{Q}}} \big(\Gamma^l_{ij}(g)\big)^* \, \mathcal{U}^{(\mathbf{Q})}(g),
    \label{eq:projection_operators}
\end{equation}
where $|\mathcal{P}_{\mathbf{Q}}|$ is the order of the group $\mathcal{P}_{\mathbf{Q}}$, $l_n$ is the dimension of the $l^{\text{th}}$ projective irreducible representation, and $\Gamma^l_{ij}(g)$ is the corresponding projective unitary representation matrix of size $l_n \times l_n$.  Using these projection operators, one can construct a symmetry-adapted basis~\cite{dresselhaus2007group,el2008symmetry,Kim1999Jun} that block-diagonalizes the BSE Hamiltonian and reduces the cost of its diagonalization, particularly at the $\Gamma$ point. It should be noted that the matrices $\Gamma^l_{ij}(g)$ in Eq.~\eqref{eq:projection_operators} for standard point groups can be obtained using packages such as {\tt Spgrep}~\cite{Shinohara2023May}. The {\tt Spgrep} package can also directly block-diagonalize a given Hamiltonian or provide a symmetry-adapted basis once the $\mathcal{U}^{(\mathbf{Q})}(g)$ matrices are provided. A detailed implementation is left for future work.

\section{Total crystal angular momentum}
Until now, we have discussed the transformation properties of excitons under crystal symmetry operations. With the definition of the $\mathscr{D}_{\mathbf{Q}}(g)$ matrices in Eqs.~\eqref{eq:exe_dmat_nonher} and \eqref{eq:exe_wf_rotate_tda}, one can obtain the irreducible representation labels of the excitonic states, which in turn allow us to derive and interpret selection rules. Although most selection rules can be deduced directly from the irreducible representations of the little point group, it is often more intuitive to consider quantum numbers analogous to total angular momentum. The concept of total angular momentum is particularly powerful, as it gives rise to conservation laws (for scattering events) and selection rules (for emission/absorption). However, in crystals, the lack of continuous rotational symmetry implies that angular momentum is not strictly conserved. This motivates the introduction of the concept of \emph{total crystal angular momentum} for excitons, similar to the concept of (linear) \emph{crystal momentum} which is related to the discrete, lattice periodic, translational invariance.

Suppose the little point group of $\mathbf{Q}$ contains an $n$-fold rotational symmetry about the axis $\hat{\mathbf{n}}$ (if multiple rotations exist, we take the largest $n$), represented by $R_n(\hat{\mathbf{n}})$. The corresponding unitary representation matrix can be written as
\begin{equation}
    \mathcal{U}^{(\mathbf{Q})}(R_n(\hat{\mathbf{n}})) 
    = \exp\left(-i\frac{2\pi}{n}J_{\hat{\mathbf{n}}}\right),
\label{eq:Umat_j_def}
\end{equation}
where $J_{\hat{\mathbf{n}}}$ is a Hermitian matrix that we refer to as the \emph{total crystal angular momentum} (or \emph{total pseudo-angular momentum}) operator along the axis $\hat{\mathbf{n}}$~\cite{PhysRevB.103.L100409}.  

Since an $n$-fold rotation satisfies $(R_n(\hat{\mathbf{n}}))^n = I$, the eigenvalues of $\mathcal{U}^{(\mathbf{Q})}(R_n(\hat{\mathbf{n}}))$ must be of the form $\exp\!\left(-i\tfrac{2\pi}{n}  j_{\hat{\mathbf{n}}}\right)$ with $j_{\hat{\mathbf{n}}} \in \{0,1,\dots,n-1\}$. Moreover, from Eq.~\eqref{eq:Q2rq_H}, we have $[\mathcal{U}^{(\mathbf{Q})}(R_n(\hat{\mathbf{n}})), \mathcal{H}^{(\mathbf{Q})}] = 0$, which implies that $J_{\hat{\mathbf{n}}}$ and $\mathcal{H}^{(\mathbf{Q})}$ are simultaneously diagonalizable. Each excitonic eigenstate of $\mathcal{H}^{(\mathbf{Q})}$ can therefore be labelled by a quantum number $j_{\hat{\mathbf{n}}}$, which we call the \emph{total crystal angular momentum} of an exciton along the ${\hat{\mathbf{n}}}$ axis.  

The $\mathscr{D}_{\mathbf{Q}}(R_n(\hat{\mathbf{n}}))$ matrices of Eq.~\eqref{eq:exe_dmat_nonher}, which are block-diagonal in degenerate subspaces, represent the symmetry operation in the excitonic basis. It follows that in the simultaneous eigenbasis of both $\mathcal{H}^{(\mathbf{Q})}$ and $\mathcal{U}^{(\mathbf{Q})}(R_n(\hat{\mathbf{n}}))$ (or equivalently $J_{\hat{\mathbf{n}}}$), these representation matrices become diagonal, with diagonal elements given by $e^{-i 2\pi j_{\hat{\mathbf{n}}}/n}$.  
 
In practice, however, the $l$-fold degenerate eigenstates $\{\Psi_S^{\mathbf{Q}}\}$ obtained by diagonalizing $\mathcal{H}^{(\mathbf{Q})}$ with eigenvalue $\varepsilon_S$ do not necessarily form a simultaneous eigenbasis of both $\mathcal{H}^{(\mathbf{Q})}$ and $J_{\hat{\mathbf{n}}}$. As a result, the representation matrix $\mathscr{D}_{\mathbf{Q};S'S}(R_n(\hat{\mathbf{n}}))$ is generally not diagonal. To obtain the simultaneous eigenstates, we diagonalize $\mathscr{D}_{\mathbf{Q};S'S}(R_n(\hat{\mathbf{n}}))$, where the resulting unitary matrix provides the transformation coefficients that rotate the arbitrary eigenstates into states that are simultaneous eigenstates of both $\mathcal{H}^{(\mathbf{Q})}$ and $J_{\hat{\mathbf{n}}}$. The resulting eigenvalues, as shown above, are $e^{-i 2\pi j_{\hat{\mathbf{n}}}/n}$, and the corresponding integers $j_{\hat{\mathbf{n}}}$ provide well-defined total crystal angular momentum quantum numbers of the excitons. It is important to emphasise that the total crystal angular momentum of the exciton does not depend on the choice of basis or on starting point, as the eigenvalues corresponding to irreducible representations remain invariant under similarity transformations, just like the character of a representation.  

We would like to point out that for nonsymmorphic space groups, where the
$n$-fold rotation corresponds to a coset representative
$g=\{R_n(\hat{\mathbf{n}})\mid\boldsymbol{\tau}\}$ with fractional
translation $\boldsymbol{\tau}$, the concept of total crystal angular
momentum can be naturally extended. This is achieved by replacing
$\mathcal{U}^{(\mathbf{Q})}$ with the redefined matrix
$\overline{\mathcal{U}}^{(\mathbf{Q})}(g)
= \mathcal{U}^{(\mathbf{Q})}(g)e^{i\mathbf{Q}\cdot\mathbf{t}}$,
and $\mathscr{D}_{\mathbf{Q}}(g)$ with
$\mathscr{D}_{\mathbf{Q}}(g)e^{i\mathbf{Q}\cdot\mathbf{t}}$, where
$\mathbf{t}
= \frac{1}{n}\sum_{i=0}^{n-1}
\left(R_n(\hat{\mathbf{n}})\right)^i \cdot \boldsymbol{\tau}$.
Since $g^n=\{I_3\mid n\mathbf{t}\}$, it follows that $\left(\overline{\mathcal{U}}^{(\mathbf{Q})}(g)\right)^n = I,$
and the eigenvalues are therefore
$e^{-i2\pi j_{\hat{\mathbf{n}}}/n}$.
If the projective representation at $\mathbf{Q}$ is trivial, then
$\mathbf{Q}\cdot\mathbf{t}=\mathbf{Q}\cdot\boldsymbol{\tau}$.
Because crystal momentum is conserved, the phase factor associated with
the fractional translation cancels exactly in the derivation of total
crystal angular-momentum conservation laws, yielding selection rules
identical to those of the symmorphic case. Accordingly, without loss of
generality, we restrict the remainder of this paper to pure rotations.

Having established how the quantum number $j_{\hat{\mathbf{n}}}$ is obtained, the next step is to understand its physical meaning. To interpret the total crystal angular momentum, it is useful to compare with the hydrogen atom. In the hydrogen atom, the solutions are atomic orbitals ($1s, 2s, 2p, \dots$) whose angular parts are spherical harmonics, labelled by the azimuthal quantum number $l$ and magnetic quantum number $m_z$. The integer $l$ specifies the irreducible representation of SO(3) to which the orbital belongs, with dimension $2l+1$, while $m_z$ represents the eigenvalue of the representation matrix (the eigenvalue is $e^{-im_z\phi}$, with $\phi$ being the rotation angle) and corresponds to the $z$-component of angular momentum (in units of $\hbar$)~\cite{Weinberg_2015}. In crystals, however, the full SO(3) rotational symmetry is broken, and the irreducible representations of SO(3) subduce as representations (generally reducible) in the crystal’s point group, lifting some degeneracies. The irreducible representations of the point group are instead labelled using notations such as the Mulliken notation ($A$, $E$, …)~\cite{dresselhaus2007group}. Thus, the irreducible representation labels obtained from the $\mathscr{D}$ matrices play the role of $l$, while the total crystal angular momentum quantum number $j$ serves the role of $m_z$ when deriving selection rules. A crucial difference is that $j_{\hat{\mathbf{n}}}$ is only defined when rotational symmetry exists in the little point group, and unlike $m_z$, it is conserved only modulo $n$ for an $n$-fold symmetry, as elaborated in Section~\ref{app:totJexe}. In this sense, total crystal angular momentum under rotations is directly analogous to linear crystal momentum under translations (which is only conserved up to a multiple of a reciprocal lattice vector).

We would like to emphasize two key points regarding the total crystal angular momentum. First, it is a well-defined quantity, regardless of whether spin is included, since it always commutes with the excitonic Hamiltonian. The $\mathscr{D}$ matrices encode the spin contributions of the underlying electronic states, which is why the label is naturally called \emph{``total"}. Moreover, the total crystal angular momentum is defined with respect to a particular axis. If multiple rotational symmetries are non-commutative, the quantum number $j_{\hat{\mathbf{n}}}$ may take different values depending on the chosen axis, and only one axis at a time admits a meaningful definition of $j_{\hat{\mathbf{n}}}$, similar to angular momentum components in the hydrogen atom.

Second, although we have introduced the concept of total crystal angular momentum for excitons, it is equally applicable to electronic states and other quasiparticles such as phonons. For electrons, the $\mathscr{D}$ matrices are replaced by the $\mathcal{D}$ matrices defined in Eq.~\eqref{eq:dmats}, which can yield half-integer $j_{\hat{\mathbf{n}}}$ values when spin is included. Similarly, for phonons, the excitonic Hamiltonian is replaced by the dynamical matrix, and the representation matrices can be computed following Appendix~\eqref{app:rotate_elph_me}. For phonons, the total crystal angular momentum is commonly referred to as the phonon pseudo-angular momentum, which is widely used in the study of circularly polarised phonons, known as chiral phonons~\cite{PhysRevLett.115.115502,Wang2024Apr}.

\subsection{Conservation of total crystal angular momentum}
\label{app:totJexe}
With the concept of total crystal angular momentum defined, we derive its conservation laws, analogous to those of total angular momentum in the hydrogen atom. These conservation laws directly govern microscopic scattering pathways involving quasiparticles such as electrons, excitons, and phonons, as discussed in detail in Section~\ref{sec:app_sec}.

Suppose there is an $n$-fold rotational symmetry along the axis $\hat{\mathbf{n}}$ in the little point group of the exciton transfer momentum $\mathbf{Q}$. Now, consider the exciton dipole matrix elements $\bra{S,\mathbf{Q}=0}  \hat{\mathbf{r}} \ket{0}$, which describe the coupling strength of the excitonic state $\ket{S,\mathbf{Q}=0}$ with photons, where $\hat{\mathbf{r}}$ is the position operator. A conservation rule for the exciton-photon matrix elements can then be derived as
\begin{equation}
\begin{aligned}
    \bra{S,\mathbf{Q}\!=\!0}  \hat{\mathbf{r}} \ket{0} 
    &= \bra{S,\mathbf{Q}\!=\!0} \hat{U}^\dagger \hat{U} \hat{\mathbf{r}} \hat{U}^\dagger \hat{U} \ket{0} \\
    &= \bra{S,\mathbf{Q}\!=\!0} \hat{\mathbf{r}} \ket{0} \, e^{-i\frac{2\pi}{n}(j_l - j_S)},
    \label{eq:tcry_J_conservation_light}
\end{aligned}
\end{equation}
where $\hat{U}$ is the unitary operator corresponding to the $n$-fold rotation, and $j_S$ and $j_l$ are the total crystal angular momenta of the exciton $S$ and the photon, respectively. This expression assumes that $\ket{S,\mathbf{Q}\!=\!0}$ is a simultaneous eigenbasis of the excitonic Hamiltonian and the unitary operator $\hat{U}$, i.e., $\hat{U}\ket{S,\mathbf{Q}\!=\!0} = e^{-i\frac{2\pi}{n}j_S}\ket{S,\mathbf{Q}\!=\!0}$. If this is not the case, one can transform to the simultaneous eigenbasis, as mentioned previously.

Equation~\eqref{eq:tcry_J_conservation_light} leads to the selection rule for the exciton-photon coupling, which is given by
\begin{equation}
    j_l - j_S = l n,
    \label{eq:Total_J_conversavtion_rule_dipoles}
\end{equation}
where $l$ is an integer. Physically, this implies that only exciton states with the same total crystal angular momentum (modulo $n$) as the photon can be optically active, similar to the case of conservation of angular momentum in the hydrogen atom.

The selection rule in Eq.~\eqref{eq:Total_J_conversavtion_rule_dipoles} can be further understood by examining how photons carry total crystal angular momentum. Under an $n$-fold rotation, the basis vectors of left- and right-circularly polarised light, $\frac{1}{\sqrt{2}}\begin{bmatrix} 1 \\ \pm i \end{bmatrix}$, transform as $e^{-i(\pm1)\frac{2\pi}{n}} \frac{1}{\sqrt{2}}\begin{bmatrix} 1 \\ \pm i \end{bmatrix}$, indicating that circularly polarised light carries total crystal angular momentum $\pm 1$ along the out-of-plane axis. This property directly links the optical selection rule to chirality: left- and right-circularly polarised light couple selectively to excitons with $j_z=-1$ and $j_z=+1$, respectively. In crystals lacking inversion symmetry, these excitons become distinguishable, and their selective light–matter coupling manifests as chiral behavior.

A similar conservation rule can also be derived for exciton-phonon interactions. Suppose the same $n$-fold rotational symmetry along $\hat{\mathbf{n}}$ exists in the little point group of the phonon crystal momentum $\mathbf{q}$, then the exciton-phonon matrix elements satisfy 
\begin{equation}
\begin{aligned}
    \bra{S',\mathbf{Q}\!+\!\mathbf{q}} \partial_{\mathbf{q}}^{\nu} V \ket{S,\mathbf{Q}} 
    &= \bra{S',\mathbf{Q}\!+\!\mathbf{q}} U^\dagger  U \ \partial_{\mathbf{q}}^{\nu} V \ U^\dagger  U \ket{S,\mathbf{Q}} \\
    &= \bra{S',\mathbf{Q}\!+\!\mathbf{q}} \partial_{\mathbf{q}}^{\nu} V \ket{S,\mathbf{Q}} \,\\ &\times 
    e^{-i\frac{2\pi}{n}(j_S+j_{\nu} - j_{S'})},
    \label{eq:tcry_J_conservation}
\end{aligned}
\end{equation}
where $j_S$, $j_\nu$, and $j_{S'}$ are the total crystal angular momenta of the initial exciton $S$, the phonon mode $\nu$, and the final exciton $S'$, respectively, and $\partial_{\mathbf{q}}^{\nu} V$ denotes the \replytoreferee{directional derivative of the self-consistent potential along the displacement vector (with dimensions of length) of the phonon mode of index $\nu$}. (See Appendix~\ref{appen:ex-ph_def} for detailed expressions for exciton-phonon matrix elements.)

According to Equation~\eqref{eq:tcry_J_conservation}, the matrix element $\bra{S',\mathbf{Q}\!+\!\mathbf{q}} \partial_{\mathbf{q}}^{\nu} V \ket{S,\mathbf{Q}}$ can be nonzero only if
\begin{equation}
    j_S + j_{\nu} - j_{S'} = l n,
    \label{eq:Total_J_conserv_rule}
\end{equation}
where $l$ is an integer\footnote{As shown in Eq.~\eqref{eq:exph_prefinal}, there is an extra term for $\mathbf{Q}=0$ corresponding to a disconnected diagram. It can be straightforwardly shown that this disconnected diagram gives the same factor as in Eq.~\eqref{eq:tcry_J_conservation}, so the conservation rule applies equally to the exciton-phonon matrix elements defined in Eq.~\eqref{eq:exph_matr_elmentQ}.}. Equations~\eqref{eq:Total_J_conserv_rule} and \eqref{eq:tcry_J_conservation} show that the total crystal angular momentum must be conserved in microscopic interactions. 

From Eq.~\eqref{eq:Total_J_conserv_rule}, we see that a completely analogous situation arises for phonons, as in the case of photons. Left- and right-circularly polarised phonon modes carry total crystal angular momentum $\pm 1$ and therefore couple selectively to other quasiparticles, such as excitons, giving rise to chiral phonon phenomena. The selective coupling among excitons, phonons, and circularly polarised light is discussed in detail in Section~\ref{sec:app_mose2}.

\section{Applications}
\label{sec:app_sec}
We now   demonstrate the application of the above methods to analyze the symmetries of excitonic states in three different materials: bulk LiF, monolayer MoSe$_2$, and bulk $h$BN. Except for LiF, we employ the TDA when obtaining excitonic energies and eigenvectors for all systems. The computational details for all calculations are provided in Appendix~\ref{section:comp_details}

\subsection{Excitons in LiF}

As a first use case, we consider excitons in bulk LiF. LiF crystallizes in a cubic lattice structure with the $Fm\bar{3}m$ space group, which does not include any non-symmorphic symmetries. It exhibits full octahedral symmetry ($O_h$ point group), consisting of 48 elements including inversion. Being non-magnetic, it also preserves time-reversal symmetry. LiF is a wide-band-gap insulator known to host strongly bound Frenkel-type excitons~\cite{PhysRevLett.81.2312}, which implies that conventional selection rules based on the Wannier type excitons  may not be applicable. This makes LiF an ideal first test case for our study.
\begin{figure}
    \centering
    \includegraphics{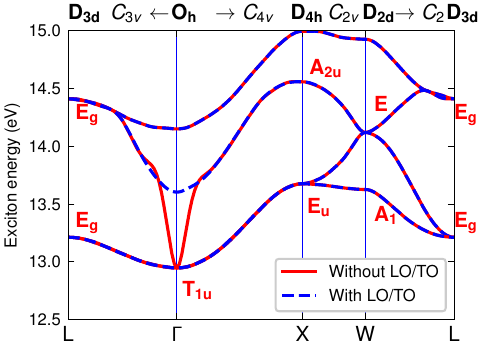}
    \caption{Exciton dispersion of LiF, computed by solving the BSE without the TDA at selected $\mathbf{Q}$ points, followed by Fourier interpolation. Excitonic states at high-symmetry points are labeled with their corresponding irreducible representation labels. The red and blue dashed curves represent the exciton dispersion with and without LO–TO splitting only at the $\Gamma$ point, respectively. The direction of the $\mathbf{Q}$ vector used for the LO–TO splitting is taken along $(1,1,1)$, which breaks the symmetries of the excitonic states at the $\Gamma$ point.}
    \label{fig:LiFdisp}
\end{figure}

To investigate the symmetries of excitons in LiF, we first analyze its excitonic dispersion. In Fig.~\ref{fig:LiFdisp}, we show the dispersion (solid red curve) of the four lowest exciton bands of LiF along high-symmetry paths, obtained from standard GW-BSE calculations. Since LiF does not possess non-symmorphic symmetries, all excitonic states at all $k$-points can be mapped to the linear representations of the crystal point group. At the $\Gamma$ point, the little point group coincides with the full $O_h$ point group. The lowest-energy exciton is triply degenerate and transforms according to the $T_{1u}$ irreducible representation. Moving away from $\Gamma$ along the $\Gamma \rightarrow L$ direction, the little point group is reduced to $C_{3v}$. Because $T_{1u}$ subduces as $A_1 \oplus E$ in $C_{3v}$, the triple degeneracy splits into a doubly degenerate $E$ band and a singly degenerate $A_1$ band. At the $L$ point, the symmetry is enlarged to $D_{3d}$, and the two lowest bands transform as $E_g$.  It is important to note that in Fig.~\ref{fig:LiFdisp}, we show the excitonic dispersion both with (blue dashed line) and without (red solid line) LO–TO splitting at the $\Gamma$ point. When the LO–TO splitting is included, the excitonic states at $\Gamma$ no longer transform according to the full point group symmetry of the crystal; instead, they follow the representations of the little point group associated with the chosen high-symmetry path direction. In the following, we focus on the excitonic dispersion computed without LO–TO splitting at the $\Gamma$ point, as it is independent of the choice of the $\mathbf{Q}$-point direction and respects full point group symmetries at the $\Gamma$ point.

Along the $\Gamma \rightarrow X$ path, the little point group is lowered to $C_{4v}$. Here, $T_{1u}$ also subduces as $A_1 \oplus E$, so the triple degeneracy again splits into a doubly degenerate $E$ band and a singly degenerate $A_1$ band. At the $X$ point, the symmetry is enlarged to $D_{4d}$, where the three lowest bands transform as $E_u$ and $A_{2u}$. Moving further along the $X \rightarrow W$ path, the symmetry is reduced to $C_{2v}$. In this case, $A_{2u}$ subduces as $B_1$, while $E_u$ subduces as $A_1 \oplus B_2$, implying that the twofold degeneracy is lifted along this path. At the $W$ point, the higher symmetry restores degeneracy where the $B_1$ and $B_2$ states merge into the $E$ irreducible representation. Finally, along the $W \rightarrow L$ path, the little point group is $C_2$, with $A_1 \rightarrow A$ and $E \rightarrow A \oplus B$. At the $L$ point, the symmetry group enlargement leads to the lowest two bands becoming degenerate again, transforming as $E_g$.  

While dispersion analysis clarifies how excitonic bands split and recombine under different little point groups, it is  insightful to look into the symmetries of excitons by examining the real-space exciton wavefunction. A conventional approach is to fix either the electron or the hole and analyze the real-space density of the other particle. For instance, to study the first exciton at the $\Gamma$ point, we fix the electron at the lithium atom located at the origin and examine the hole density. This choice for the electron position is possible because the Bloch state of the conduction band minimum possesses a strong Li $2s$ admixture. The electron thus has an enhanced probability density at the position of a Li nucleus. With the electron localized in a highly symmetric position, the symmetry of the exciton can then be effectively determined by that of the hole part.

\begin{figure*}
    \centering
    \includegraphics[width=\linewidth]{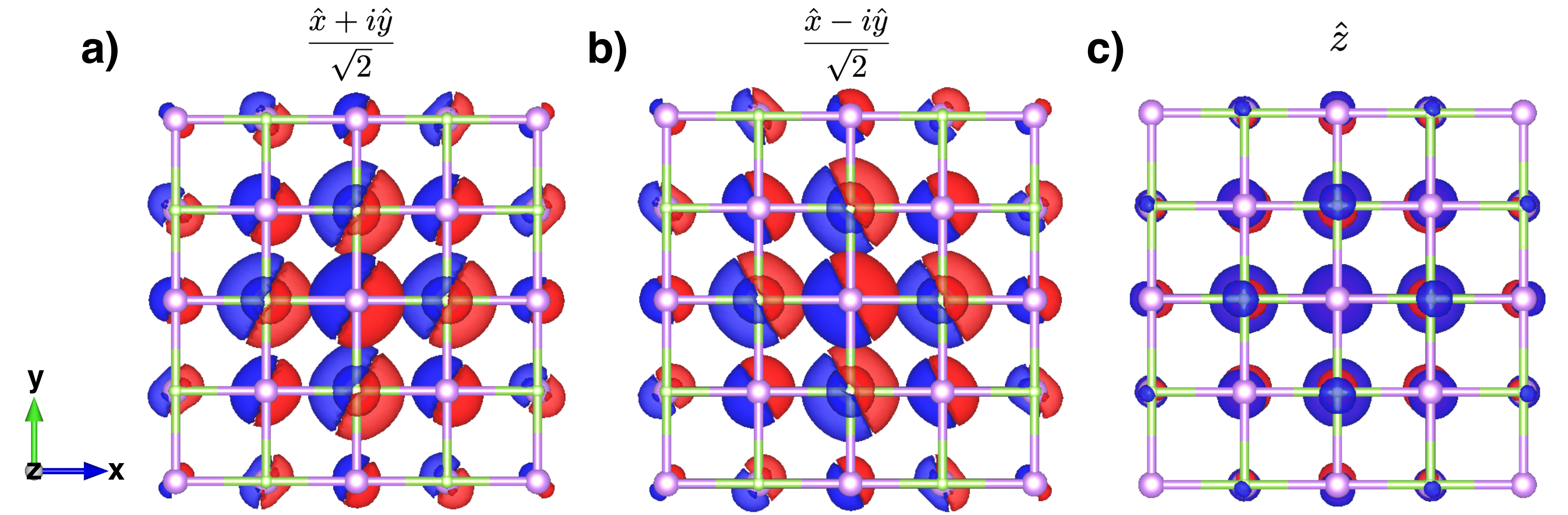}
    \caption{Real-space hole density, multiplied by the sign of the wavefunction as defined in Eq.~\eqref{eq:hole_den_ex}, for the first triply degenerate excitons of LiF at the $\Gamma$ point, with the electron fixed at the center of the Li atom located at the origin. Panels (a), (b), and (c) correspond to the three degenerate excitons, each selectively coupling to light polarized along the $\frac{\hat{x}+i\hat{y}}{\sqrt{2}}$, $\frac{\hat{x}-i\hat{y}}{\sqrt{2}}$, and $\hat{z}$ directions, respectively. The density is shown from the top view along the $z$ axis. Pink and green spheres denote lithium and fluorine atoms, while red and blue indicate the negative and positive lobes of the wavefunction.}
    \label{fig:LiF_gamma_exciton_holeDensity}
\end{figure*}

In Fig.~\ref{fig:LiF_gamma_exciton_holeDensity}, we depict the hole density of the first triply degenerate exciton at the $\Gamma$ point with the electron fixed at the origin. As shown in the figure, the hole density is predominantly composed of fluorine $p_z$ orbitals. To retain phase information, the density is multiplied by the sign of the real part of the wavefunction, i.e.,
\begin{equation}
\rho^{S}_{\text{hole}}(\mathbf{r}_h)
= \text{sgn}\!\left(\text{Re}\, \Psi_S(\mathbf{r}_e\!=\!0,\mathbf{r}_h) \right)
\cdot \big|\Psi_S(\mathbf{r}_e\!=\!0,\mathbf{r}_h)\big|^2,
\label{eq:hole_den_ex}
\end{equation}
Here, $\text{sgn}$ and Re denote the sign function and the real part, respectively, and $\mathbf{r}_h$ denotes the hole position. This procedure allows us to visualize both the amplitude and the phase structure of the excitonic wavefunction. In subplots \ref{fig:LiF_gamma_exciton_holeDensity}(a)–(c), we show the hole densities of the lowest three triply degenerate excitons at the $\Gamma$ point, viewed from the top along the $z$ axis. These states are simultaneously eigenstates of the $C_4$ rotation operator about the $z$ axis, with total crystal angular momentum $j=+1,-1,0$, and therefore they couple selectively to light polarized along $\tfrac{\hat{x}+i\hat{y}}{\sqrt{2}}$, $\tfrac{\hat{x}-i\hat{y}}{\sqrt{2}}$, and $\hat{z}$, respectively.

As shown in the subplots, with respect to the $xy$ horizontal mirror plane, the excitons in panels \ref{fig:LiF_gamma_exciton_holeDensity}a and \ref{fig:LiF_gamma_exciton_holeDensity}b are symmetric, while the exciton in panel \ref{fig:LiF_gamma_exciton_holeDensity}c is antisymmetric. This yields a character of $1$ for the horizontal mirror symmetry operation, consistent with the $O_h$ character table. Likewise, under twofold rotations about the $\hat{z}$ axis, the excitons in panels \ref{fig:LiF_gamma_exciton_holeDensity}a and \ref{fig:LiF_gamma_exciton_holeDensity}b are odd, whereas the exciton in \ref{fig:LiF_gamma_exciton_holeDensity}c is even, giving a character of $-1$ for the three $C_2$ rotations, again in agreement with group theory. In a similar way, one can determine the characters of the remaining operations as well. However, this procedure requires substantial manual effort\cite{Paleari_2018}, and the ambiguity in choosing the electron/hole position makes it less robust than the method presented in this paper.

\begin{figure}
    \centering
    \includegraphics{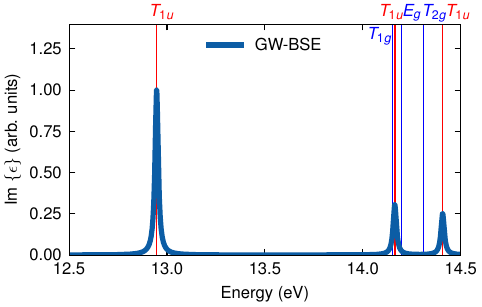}
    \caption{Absorption spectrum of LiF computed by solving the Bethe--Salpeter equation (blue solid line). The red and blue vertical lines indicate the positions of optically bright and dark excitons, respectively. The labels correspond to the irreducible representation labels of excitons.}
    \label{fig:LiFexe}
\end{figure}

Having characterized excitonic dispersions and real-space symmetries, we now turn to their role in optical absorption spectroscopy. Since only excitonic states close to the $\Gamma$ point can couple directly to light, we restrict our focus to the states at the $\Gamma$ point. The optical absorption spectrum is obtained from the imaginary part of the dielectric tensor as a function of photon energy along the polarization direction $\mathbf{e}^\mu$, which is given by~\cite{PhysRevB.62.4927,RevModPhys.74.601}
\begin{equation}
    \text{Im}\{\epsilon^{\mu\mu}(\omega)\} \propto \sum_{S} \rho_S^{L;\mu} \, \rho_S^{R;\mu} \, \delta(\omega-E_S),
     \label{eq:eps_abs}
\end{equation}
where $\rho^{L;\mu}_S = \bra{S^L} \hat{\mathbf{r}}\cdot \mathbf{e}^\mu \ket{0}$ and $\rho^{R;\mu}_S = \bra{0} \hat{\mathbf{r}}\cdot \mathbf{e}^\mu \ket{S^R}$ are the exciton dipole matrix elements, the superscripts $L/R$ denote left and right eigenstates, and $\hat{\mathbf{r}}$ is the position operator.  Within TDA, $|S^R\rangle = |S^L\rangle$, which implies that $\rho_S^L = (\rho_S^R)^*$. We performed all calculations on LiF without use of the TDA in order to confirm that our procedure of symmetry assignment works both within and without the TDA. For LiF, the results with and without the TDA are practically indistinguishable as expected from the presence of the large band gap.

In Fig.~\ref{fig:LiFexe}, we present the optical absorption spectrum computed using \emph{ab initio} methods. The red and blue vertical lines indicate bright and dark excitons, respectively. Since absorption is directly proportional to the exciton dipole matrix elements, only excitons that transform like dipole operators are optically active and appear in the spectrum. Although many excitonic states exist, only a few are bright, as dictated by their underlying symmetries.  

To understand the dipole selection rule, we compute the irreducible representation labels of the excitonic states, as shown in Fig.~\ref{fig:LiFexe}. The lowest bright exciton, which is triply degenerate, transforms according to the $T_{1u}$ irreducible representation, while the first dark exciton, also triply degenerate, transforms according to $T_{1g}$. Higher-energy dark excitons belong to irreducible representations other than $T_{1u}$. Since the dipole operators transform as $T_{1u}$ in the $O_h$ point group, only excitons of this symmetry can couple to light, thereby confirming the validity of our method.  

\subsection{Resonant Raman scattering in Monolayer MoSe$_2$}
\label{sec:app_mose2}
For the second test case, we investigate the symmetries of excitonic states in monolayer MoSe$_2$. Monolayer MoSe$_2$ crystallizes in a two-dimensional hexagonal lattice with space group $P\bar{6}m2$. Its point group is $D_{3h}$ which consists of 12 symmetry operations without the inversion symmetry. In contrast to LiF, MoSe$_2$ exhibits strong spin–orbit coupling, which requires the electronic states to be described using two-component spinors. 

\begin{figure}
    \centering
    \includegraphics{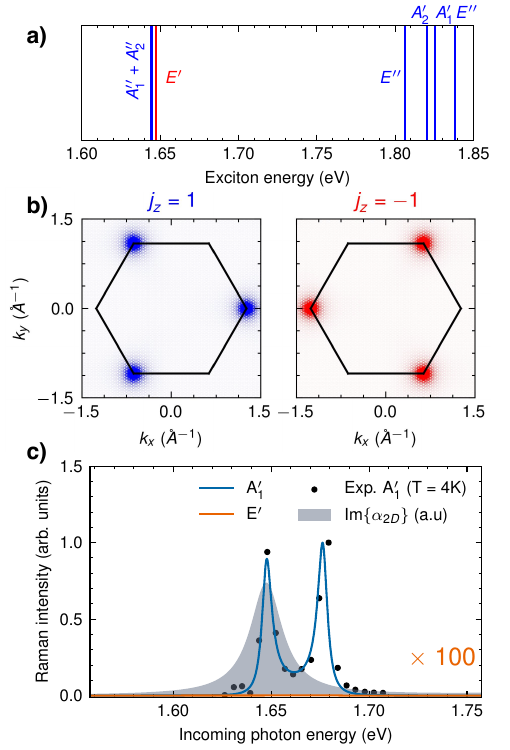}
    \caption{
    Excitons in monolayer MoSe$_2$.  
(a) Exciton energy spectrum at the $\Gamma$ point with corresponding irreducible representations. The blue/red vertical lines indicate the positions of dark/bright excitons, respectively.  
(b) Reciprocal space plot of the $\mathscr{A}_{1s}$ exciton wavefunction in the simultaneous eigenstate of the total crystal angular momentum matrix and the exciton Hamiltonian.  
(c) Resonant Raman spectrum as a function of incoming photon energy. The $A_1'$ and $E'$ Raman modes are represented by blue and orange lines, respectively. The grey shading corresponds to the imaginary part of the 2D polarizability tensor, representing the absorption spectrum. The black dots denote experimental data for the $A_1'$ mode, taken from Ref.~\cite{McDonnell_2020}. The absorption spectrum, and Raman spectrum are rigidly shifted in energy such that the first peak of the experimental Raman peak in (c) is aligned. 
}
    \label{fig:mose2}
\end{figure}

First, we examine the zero-momentum excitonic states of monolayer MoSe$_2$. At the center of the Brillouin zone, the little point group coincides with the full point group of the crystal. As a result, the zero-momentum excitonic states can be classified according to the irreducible representations of the $D_{3h}$ point group.

   In Fig.~\ref{fig:mose2}a, we present the energies of the first few zero-momentum excitonic states, calculated using the GW-BSE formalism. The states are labeled by their irreducible representations, with in-plane bright and dark excitons shown in red and blue, respectively. The lowest exciton, which is doubly degenerate due to time-reversal symmetry, is spin-forbidden and therefore optically dark for in-plane light polarization, as it transforms according to the $A''_1 \oplus A''_2$ representation. Since the out-of-plane dipole operator $\hat{z}$ transforms as $A''_2$, this exciton can possess a finite dipole moment along the out-of-plane direction (although, this being a thin 2D system, it may be unrealistic to consider a purely $\hat{z}$-polarized incoming field). By contrast, dark excitons around $\sim 1.8$ eV do not contain the $A''_2$ component and therefore remain strictly dipole forbidden. In comparison, the lowest in-plane bright exciton, commonly referred to as the $\mathscr{A}_{1s}$ exciton, transforms as the $E'$ representation and is doubly degenerate. Because the in-plane dipole operators $\hat{x},\hat{y}$ also transform as $E'$, this exciton possesses a finite in-plane dipole moment, making it optically active for in-plane polarizations.

In recent years, there has been significant interest in these first bright excitons (also in other monolayer transition-metal dichalcogenides (TMDCs)), which have been shown to exhibit chirality by selectively coupling to left- and right-circularly polarised light~\cite{He2014Jul,RevModPhys.90.021001}. Moreover, one of the remarkable properties of these materials is their ability to selectively populate the electron and hole densities in the inequivalent $K$ and $K'$ valleys using left- or right-circularly polarised light. A natural question then arises: where does this chirality originate? To understand this phenomenon, we examine the total crystal angular momentum. 

Due to the presence of threefold rotational symmetry in the little point groups of the $\Gamma$ and $K^{(')}$ points, the total crystal angular momenta of excitons or phonons are restricted to the set $\{-1,0,1\}$ (with $-1 \equiv 2$, since the total crystal angular momentum is defined modulo $n$ for an $n$-fold rotation). States transforming according to the $E$ representation, which are singly or doubly degenerate, therefore carry total crystal angular momenta of $\pm 1$. Consequently, simultaneous eigenstates of excitons or phonons transforming under $E$ modes can selectively couple to other quasiparticles, provided that the interaction conserves total crystal angular momentum.

For the $\mathscr{A}_{1s}$ exciton, the total crystal angular momenta along the principal axis are $j_z = \pm 1$. These values can also be obtained by diagonalising the two-dimensional excitonic representation matrix $\mathscr{D}_{\mathbf{0};S'S}(R_3(\hat{\mathbf{z}}))$ corresponding to the threefold rotation $R_3(\hat{\mathbf{z}})$. Since circularly polarised light carries a total crystal angular momentum of $\pm 1$ along the out-of-plane direction, the $\mathscr{A}_{1s}$ exciton collapses onto one of the simultaneous eigenstates upon absorption or emission of left- or right-circularly polarised light, in accordance with the selection rule given in Eq.~\eqref{eq:Total_J_conversavtion_rule_dipoles}.

In Fig.~\ref{fig:mose2}b, we show the exciton density in reciprocal space, defined as $\sum_{v,c}|X^{S;(\mathbf{Q})}_{\mathbf{k}cv}|^2$, for the $\mathscr{A}_{1s}$ excitons, which simultaneously diagonalise the $C_3$ rotation operator. Owing to strong spin–orbit coupling and broken inversion symmetry, we find that for each $j_z$, the exciton is strongly localised in the corresponding $K$ or $K'$ valley. This localisation of the excitonic density at the $K/K'$ valley enables the selective excitation of electron and hole densities in a specific valley using circularly polarised light, thereby inducing valley polarisation.

Despite the rigorous valley-dependent selection rules, the experimentally measured degree of valley polarisation rarely reaches $100\%$. This valley depolarisation is attributed to dephasing mechanisms such as exciton–phonon scattering~\cite{Molina-Sanchez2017Aug}. Exciton–phonon interactions not only limit valley polarisation but also crucially determine the intrinsic properties of excitons, such as their lifetimes~\cite{Chan2023May}. Furthermore, these interactions are central to a variety of optical scattering phenomena such as resonant Raman scattering~\cite{doi:10.1126/sciadv.abb5915}.

To investigate the selection rules that govern exciton-phonon scattering processes, we focus on phonon-mediated Stokes resonant Raman scattering in monolayer MoSe$_2$. Following the approach of Refs.~\cite{PhysRevB.99.174312,doi:10.1126/sciadv.abb5915}, we compute the Raman intensities at zero temperature within the Tamm–Dancoff approximation~\cite{tda_ref}:
{\small \begin{equation}
    \label{eq:tda_ram11}
    \begin{split}
        I^\lambda \propto \frac{\omega_{\mathrm{L}}-\omega_{\lambda}}{\omega_{\mathrm{L}}} 
        \Bigg| \sum_{S, S^{\prime}} \frac{ \left(d^\nu_{S}\right)^* \left(\mathcal{G}_{S^\prime,S}^\lambda(\mathbf{0},\mathbf{0})\right)^* d_{S^\prime}^\mu }{\left(\hbar \omega_{\mathrm{L}}-E_{S^\prime}+i \gamma\right)\left(\hbar \omega_{\mathrm{L}}-\hbar \omega_\lambda-E_{S}+i \gamma\right)}  \\
         + \sum_{S, S^{\prime}} \frac{\left(d_{S^\prime}^\mu\right)^* \mathcal{G}_{S^\prime,S}^\lambda (\mathbf{0},\mathbf{0})d^\nu_{S}}{\left(\hbar \omega_{\mathrm{L}}+E_{S^\prime}-i \gamma\right)\left(\hbar \omega_{\mathrm{L}}-\hbar \omega_\lambda+E_{S}-i \gamma\right)} \Bigg|^2,
    \end{split}
\end{equation}}
where $\mu/\nu$ denote the polarisations of the incident and scattered light, $\omega_{\mathrm{L}}$ is the incoming photon energy, and $\omega_{\lambda}$ is the phonon frequency of mode $\lambda$. The summation is over excitonic states $S(S')$ with energies $E_{S(S')}$ and decay constant $\gamma$. The velocity matrix element $d^{\mu}_S = \bra{S} \hat{v}^\mu \ket{0}$ describes photon absorption into the excitonic state $|S\rangle$ along polarisation $\mu$, while $(\mathcal{G}^{\lambda}_{S',S}(\mathbf{0},\mathbf{0}))^{*}$ denotes the exciton–phonon coupling for scattering from $S$ with crystal momentum $\mathbf{Q}=\mathbf{0}$ to $S'$ with the same crystal momentum via the emission of a phonon of mode $\lambda$ and momentum $\mathbf{q}=\mathbf{0}$ (see Eq.~\eqref{eq:exph_matr_elmentQ} for details).

In Fig.~\ref{fig:mose2}c, we show the calculated resonant Raman intensities (solid lines) for the $A'_1$ (blue) and $E'$ (orange) phonon modes as a function of the incoming photon energy near the $\mathscr{A}_{1s}$ exciton. Experimental Raman intensities for the $A'_1$ mode, measured at $T=4\ \text{K}$ and taken from Ref.~\cite{McDonnell_2020}, are plotted as black dots, while the grey shading denotes the imaginary part of the in-plane polarisability (absorption spectrum). As shown in the figure, near the optical gap the absorption spectrum is dominated by the $\mathscr{A}_{1s}$ exciton.  

Moreover, the $A'_1$ intensity profile exhibits two pronounced resonances at 1.65 eV and 1.68 eV. The first, at the $\mathscr{A}_{1s}$ exciton energy, corresponds to the direct resonant excitation of the $\mathscr{A}_{1s}$ exciton followed by scattering through the $A'_1$ phonon. The second resonance, shifted upward by one $A'_1$ phonon energy, has no counterpart in the absorption spectrum and arises from phonon-assisted recombination of the $\mathscr{A}_{1s}$ exciton.  

The most striking observation in Fig.~\ref{fig:mose2}c is the enormous difference in Raman intensity between the $A'_1$ and $E'$ modes. To understand this difference, we first note that, from Eq.~\eqref{eq:tda_ram11}, the numerator of the Raman matrix element is finite only if both the incoming ($S$) and outgoing ($S'$) excitonic states are dipole active. Thus, only bright excitons can participate in one-phonon resonant Raman scattering. Since the $\mathscr{A}_{1s}$ exciton is the only bright state near the optical gap, the dominant scattering pathway is $\mathscr{A}_{1s} \!\to\! \mathscr{A}_{1s}$ via a phonon.  Therefore, to understand the difference in Raman intensities between the $A'_1$ and $E'$ modes, we need to analyse this scattering pathway, which is described by the exciton–phonon matrix element $\mathcal{G}_{S',S}^\lambda(\mathbf{0},\mathbf{0})$ with $S=S'=\mathscr{A}_{1s}$. This matrix element represents the probability amplitude of the scattering process.

If we analyze the selection rules in terms of Raman-active modes using irreducible representations, we trivially conclude that in order for $\mathcal{G}_{S',S}^\lambda(\mathbf{0},\mathbf{0})$ to be finite, both the $A_1^\prime$ and $E^\prime $ phonon modes \textit{can} work: $A_1^\prime$ simply because it is the trivial representation (so $E^\prime \otimes A_1^\prime = E^\prime$) and $E^\prime $ because the same representation is contained in $E^\prime \otimes E^\prime = A_1^\prime \oplus A_2^\prime \oplus E^\prime$. However, this does not explain why the latter mode is not resonantly enhanced.

To understand this behavior, we examine the role of total crystal angular momentum. Owing to the threefold rotational symmetry of the little point  group at the $\Gamma$ point, both excitons and phonons must conserve total crystal angular momentum upto modulo 3 for the exciton–phonon coupling matrix element $\mathcal{G}_{S',S}^\lambda(\mathbf{0},\mathbf{0})$ to be nonzero, which can be expressed as
\begin{equation}
    j_{z;S'} = j_{z;S} + j_{z;\lambda} + 3l,
\label{eq:conseJmain}
\end{equation}
where $j_{z;S}$ and $j_{z;S'}$ denote the total crystal angular momenta of the initial and final excitonic states, $j_{z;\lambda}$ represents that of the phonon, and $l$ is an integer. 

For the $A'_1$ phonon, $j_{z;\lambda}=0$, implying that the total crystal angular momentum must remain unchanged ($j_{z;S'}=j_{z;S}$) for a finite $\mathcal{G}_{S',S}^\lambda(\mathbf{0},\mathbf{0})$. By contrast, the $E'$ phonon carries $j_{z;\lambda}=\pm 1$, requiring a change of $\pm 1$ in the total exciton angular momentum. Consequently, we gain the information that the $A'_1$ phonon allows only intra-valley scattering, while the $E'$ phonon permits only inter-valley scattering between $\mathscr{A}_{1s}$ excitons. However, intervalley scattering is strongly suppressed in monolayer MoSe$_2$ due to the negligible overlap between excitonic wavefunctions localized in opposite valleys (as shown in Fig.~\ref{fig:mose2}b) and strong spin-selection rules, whereas intra-valley scattering is unrestricted. As a result, the $E'$ mode exhibits resonant Raman intensities that are orders of magnitude less than those of the $A'_1$ mode, reflecting the weaker scattering channel and explaining the observed behavior in Fig.~\ref{fig:mose2}c.

\subsection{Phonon-assisted luminescence of \emph{h}BN}
For the third test case, we look into excitons in bulk $h$BN which crystallizes in a hexagonal lattice with space group $P6_3/mmc$, including nonsymmorphic symmetries. Its point group is $D_{6h}$, containing 24 elements including inversion. $h$BN is a wide-band-gap insulator that hosts strongly bound excitons with a mixed Wannier–Frenkel character~\cite{PhysRevB.94.125303}.
\begin{figure*}
    \centering
\includegraphics[width=\linewidth]{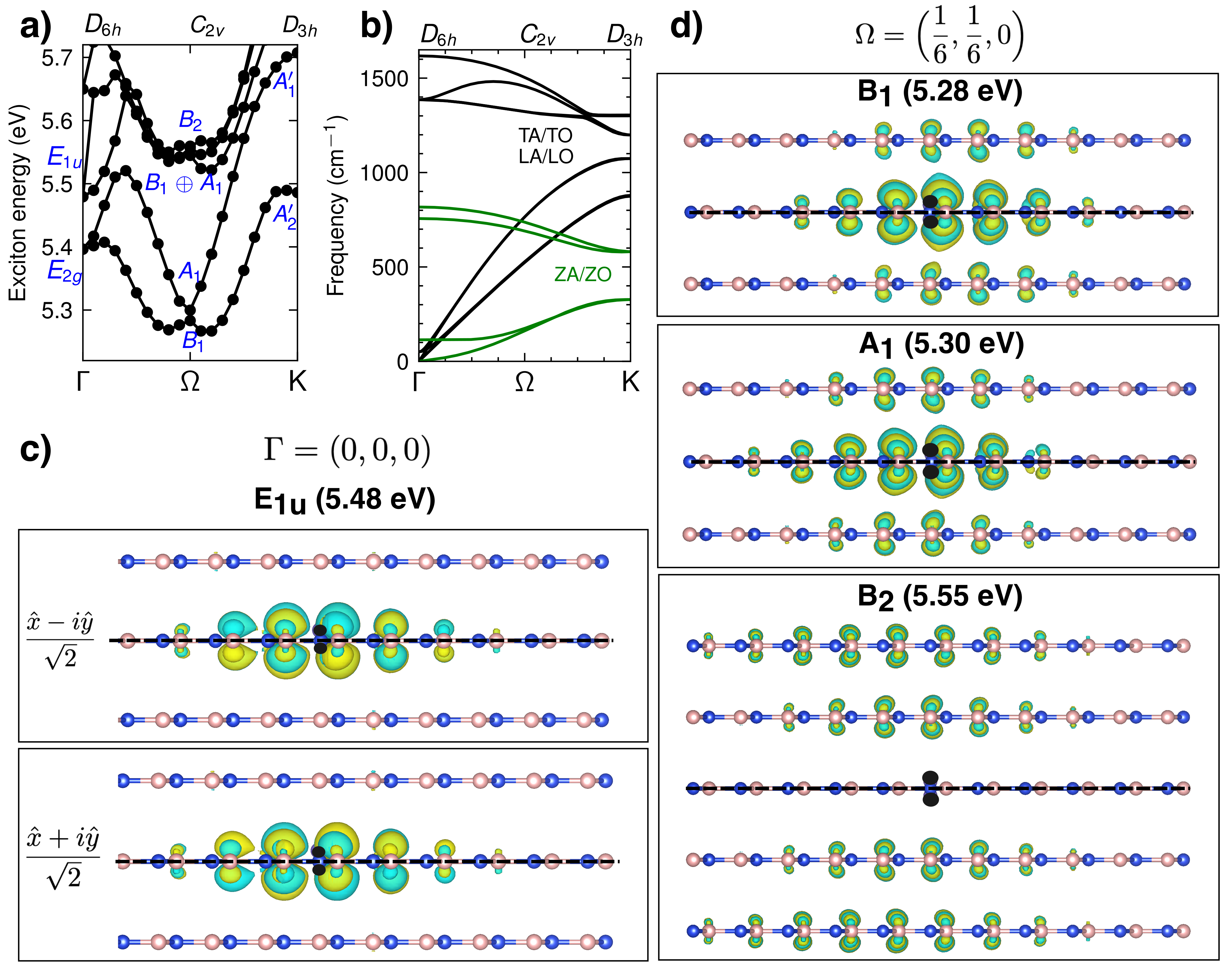}
    \caption{Excitons in $h$BN.
(a) Exciton dispersion of $h$BN computed using GW-BSE. Black dots mark the $\mathbf{Q}$ points where calculations were performed. Excitonic states at high-symmetry points are labeled with their corresponding irreducible representations.
(b) Phonon dispersion of $h$BN. Black lines denote in-plane phonon modes that are even under horizontal mirror symmetry, while green lines denote out-of-plane phonon modes that are odd with respect to the horizontal mirror plane.
(c) Electronic density of excitons, as defined in Eq.~\eqref{eq:elec_den_ex}, at the $\Gamma$ point $\left(0,0,0\right)$ for the first two in-plane bright excitons. These excitons are degenerate, symmetric under the horizontal mirror plane, and selectively couple to left- ($\tfrac{\hat{x}-i\hat{y}}{\sqrt{2}}$) and right-circularly ($\tfrac{\hat{x}+i\hat{y}}{\sqrt{2}}$) polarized light. The holes are fixed near a nitrogen atom and appear as mirror images of each other.
(d) Electronic density of the first, second, and fifth lowest-energy excitons at $\Omega = \left(\tfrac{1}{6}, \tfrac{1}{6}, 0\right)$. In (c) and (d), the dashed horizontal line indicates the mirror plane.}
    \label{fig:hBN_disp_wf}
\end{figure*}

To investigate the symmetries of excitons in $h$BN, we analyze its excitonic dispersion. In Fig.~\ref{fig:hBN_disp_wf}a, we show the five lowest-energy exciton bands along the $\Gamma\rightarrow K$ high-symmetry path, obtained from standard GW-BSE calculations. In addition, Fig.~\ref{fig:hBN_disp_wf}b shows the phonon dispersion of $h$BN, obtained from DFPT calculations. For all the $\mathbf{Q}$ points along $\Gamma\rightarrow K$, the condition $\mathbf{G}_0\!=\!0$ in Eq.~\eqref{eq:new_rep_group_mul} holds, so all excitonic states and phonon modes at these $\mathbf{Q}$-points can be mapped onto linear representations of the crystal point group. At the $\Gamma$ point, the little point group coincides with the full $D_{6h}$ point group. The lowest-energy exciton at the $\Gamma$ point is doubly degenerate and transforms according to the $E_{2g}$ irreducible representation. Therefore, it is even under inversion, twofold rotation about the principal axis, and the horizontal mirror symmetry. In contrast, the next-lowest exciton at $\Gamma$ transforms as $E_{1u}$, making it odd under inversion and the twofold rotation about the principal axis, while remaining even under the horizontal mirror symmetry. Since the in-plane dipole operators transform as the $E_{1u}$ representations, this exciton is dipole-allowed and therefore optically bright.

When moving along the $\Gamma \rightarrow K$ path, the symmetry is reduced to $C_{2v}$, with the horizontal mirror plane as one of the symmetry elements of the little point group. Under this reduction, the $E_{2g}$ representation of the $D_{6h}$ point group subduces as $A_1 \oplus B_1$ in $C_{2v}$, causing the $E_{2g}$ exciton to split into two nondegenerate states transforming as $A_1$ and $B_1$. At the $K$ point, where the symmetry increases to $D_{3h}$, the two lowest excitons transform as the $A_1'$ and $A_2'$ irreducible representations. It is important to highlight that the two lowest branches along the $\Gamma \rightarrow K$ path remain even under the horizontal mirror symmetry.

To gain visual insight into the horizontal mirror symmetry of excitons along the $\Gamma \rightarrow K$ path, we visualize the exciton wavefunctions in real space, following a strategy similar to that used for LiF. In $h$BN, the lowest excitons are primarily composed of excitations from nitrogen $p_z$ orbitals to boron $p_z$ orbitals ~\cite{PhysRevB.94.125303}, suggesting the nitrogen $p_z$ orbital as a natural choice for fixing the hole. Since the $p_z$ orbital has a nodal plane, we cannot fix the hole at the position of a nitrogen nucleus, but we can position it slightly above. However, this choice alone breaks the horizontal mirror symmetry, as the mirror operation does not map the hole back onto itself. To restore the symmetry, we adopt the approach of Ref.~\cite{Paleari_2018}, fixing two hole positions symmetric with respect to the horizontal mirror plane: one slightly above and the other at the same distance below the nitrogen atom. These two positions are mirror images of each other. Using this construction, we define the antisymmetrized wavefunction
\begin{equation}
    \tilde{\Psi}(\mathbf{r}_e) = \Psi(\mathbf{r}_e, z \!+\! z_p) - \Psi(\mathbf{r}_e, z \!-\! z_p),
    \label{eq:psi_anti_symm_hBn}
\end{equation}
where $\Psi(\mathbf{r}_e, z \pm z_p)$ is the excitonic wavefunction with the hole fixed at a vertical displacement $\pm z_p$ from the mirror plane, and $\mathbf{r}_e$ is the electron position. Since nitrogen $p_z$ orbitals are odd under horizontal mirror symmetry, a symmetric combination would nearly cancel the wavefunction. The antisymmetric construction of Eq.~\eqref{eq:psi_anti_symm_hBn} therefore ensures that $\tilde{\Psi}(\mathbf{r})$ is symmetric for antisymmetric excitons and antisymmetric for symmetric excitons. Similar to Eq.~\eqref{eq:hole_den_ex}, we then define the electronic density as
\begin{equation}
    \rho^{\text{exc}}_{\text{elec}}(\mathbf{r}_e) = \text{sgn}(\text{Re}\{\tilde{\Psi}(\mathbf{r}_e)\}) \cdot |\tilde{\Psi}(\mathbf{r}_e)|^2.
    \label{eq:elec_den_ex}
\end{equation}

Figures~\ref{fig:hBN_disp_wf}c and \ref{fig:hBN_disp_wf}d show the electronic densities obtained from Eq.~\eqref{eq:elec_den_ex}, with two holes fixed near a nitrogen atom (indicated by black circles). In Fig.~\ref{fig:hBN_disp_wf}c, we show the second-lowest excitons at the $\Gamma$ point, which belong to the $E_{1u}$ representation and couple selectively to left- and right-circularly polarized light, i.e., they are simultaneously eigenstates of the $C_3$ rotation operator.
 In Fig.~\ref{fig:hBN_disp_wf}d, we show the exciton wavefunctions at the $\Omega = \left(\tfrac{1}{6}, \tfrac{1}{6}, 0\right)$ point for the two lowest excitons, which  transform as $B_1$ and $A_1$, and for the fifth lowest exciton, which transforms as $B_2$.

These real-space visualizations provide direct confirmation of the symmetries of the excitons. The $E_{1u}$ excitons at $\Gamma$ are symmetric under the horizontal mirror plane, which results in antisymmetric electron densities, as seen in Fig.~\ref{fig:hBN_disp_wf}c. A similar correspondence holds at the $\Omega$ point, where symmetric exciton wavefunctions yield antisymmetric electron densities. The most striking case is the antisymmetric $B_2$ exciton, where the absence of electron density in the layer closest to the fixed holes follows directly from symmetry. Because boron $p_z$ orbitals are odd under mirror reflection, any electron density in the nearest layer would make the overall exciton wavefunction symmetric, which is forbidden.

\begin{figure}[!ht]
    \centering
    \includegraphics{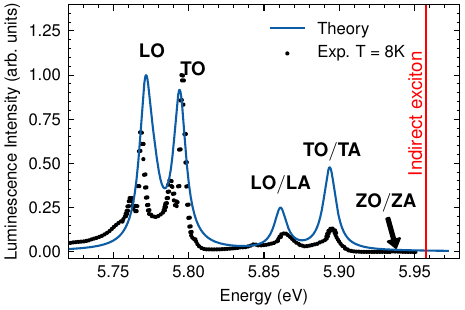}
    \caption{Phonon-assisted luminescence spectrum of $h$BN. Experimental data (dots) are taken from Ref.~\cite{Zanfrognini2023Nov}, while the computed spectrum is shown as a solid blue line. The computed luminescence is rigidly shifted in energy so that its first peak aligns with the first peak of the experimental spectrum.}
    \label{fig:hbn_lumin}
\end{figure}
Having established the even- and odd-parity of excitonic states under horizontal mirror symmetry at $\Omega$ and $\Gamma$, we now turn to its direct implications on the phonon-assisted luminescence of $h$BN~\cite{Zanfrognini2023Nov,PhysRevLett.122.187401,Cassabois2016}. In particular, we highlight the role of horizontal mirror symmetry on the exciton–phonon matrix elements, which become evident in the luminescence spectrum. To compute phonon-assisted luminescence, we follow the approach of Ref.~\cite{Zanfrognini2023Nov}, expressed as
\begin{equation}
\begin{aligned}
    I \propto \!  \sum_{S',\mathbf{Q},\lambda,\mu} &  \left\{
   \frac{ e^{-\frac{E_{S'}^{\mathbf{Q}} - E^{\vphantom{\mathbf{Q}}}_m}{k_B T}} \, (1 + n_{\lambda, \mathbf{Q}}) \, 
   \delta\!\left(\omega^{\vphantom{\mathbf{Q}}}_L - E^\mathbf{Q}_{S'}+ \hbar \omega^{\vphantom{\mathbf{Q}}}_{\lambda, \mathbf{Q}}\right)}
   {E^\mathbf{Q}_{S'} - \hbar \omega^{\vphantom{\mathbf{Q}}}_{\lambda, \mathbf{Q}}} \right. \\ &
   \left. \times \left| \sum_{S} 
   \frac{ \left(d^\mu_{S} \, \mathcal{G}^{\vphantom{*}}_{S',S}(\mathbf{0},\mathbf{Q})\right)^*}
   {E^{\vphantom{\mathbf{Q}}}_{S} -E^\mathbf{Q}_{S'}+ \hbar \omega^{\vphantom{\mathbf{Q}}}_{\lambda, \mathbf{Q}}}  
   \right|^2 \right\} ,
\end{aligned}
    \label{eq:ph_assited_lumin_expre}
\end{equation}
where $E_m$ is the lowest exciton energy, $T$ is the exciton temperature, $k_B$ is the Boltzmann constant, $E_S^\mathbf{Q}$ denotes the exciton energy at $\mathbf{Q}$, $n_{\lambda,\mathbf{Q}}$ is the Bose factor for a phonon of mode $\lambda$ with crystal momentum $\mathbf{Q}$, and the remaining indices follow the conventions of Eq.~\eqref{eq:tda_ram11}.  

In Fig.~\ref{fig:hbn_lumin}, we show the experimental and computed phonon-assisted luminescence spectrum of $h$BN. The dominant phonons contributing to the transition matrix element in Eq.~\eqref{eq:ph_assited_lumin_expre} are located near the midpoint between $\Gamma$ and $K$. This is due to the Boltzmann factor in Eq.~\eqref{eq:ph_assited_lumin_expre}, which exponentially suppresses the matrix elements as the exciton energy moves away from the minimum exciton energy $E_m$. As seen in Fig.~\ref{fig:hBN_disp_wf}a, the exciton minimum occurs near the  $\Omega$ point, which implies that most of the finite-momentum excitonic population resides near the $\Omega$ point at thermal equilibrium. Consequently, the dominant scattering pathway involves excitons near $\Omega$ (initial state) being scattered by phonons near $\Omega$ into excitons at the $\Gamma$ point (final state).

The most striking observation in Fig.~\ref{fig:hbn_lumin} is that only in-plane phonon modes at the $\Omega$ point contribute to the luminescence spectrum. This follows directly from symmetry-based selection rules. Similar to the Raman case, Eq.~\eqref{eq:ph_assited_lumin_expre} requires the outgoing excitons (indexed by $S$) to be bright in order to yield a finite contribution. This means that the final excitonic state at the $\Gamma$ point must transform as the $E_{1u}$ representation (we restrict ourselves to in-plane dipoles, since excitons with out-of-plane dipoles lie higher in energy and possess very weak dipole strength~\cite{Paleari_2018,PhysRevB.94.125303,PhysRevLett.96.126104,PhysRevLett.96.026402}). Since both the initial and final excitonic states are even under horizontal mirror symmetry, the exciton–phonon matrix elements are finite only for phonon modes that are also even. As a result, out-of-plane phonons, which are odd under this symmetry, cannot couple efficiently to the lowest-energy excitonic scattering states and therefore contribute negligible spectral weight. In contrast, in-plane phonon modes are even and thus dominate the observed luminescence.  

It is important to note that this symmetry selection rule serves as a clear fingerprint of the hexagonal phase. By contrast, rhombohedral stacking lacks horizontal mirror symmetry, and consequently all phonon modes contribute to the luminescence spectrum~\cite{Zanfrognini2023Nov}.

\section{Conclusion}
In conclusion, this work establishes a rigorous framework for analyzing the symmetries of excitonic states in crystalline materials. We show that excitons can be classified according to the projective representations of the little point group of $\mathbf{Q}$ and provide an \emph{ab initio} approach to assign irreducible-representation labels without requiring real-space wavefunction analysis. We also introduce the concept of total crystal angular momentum in the presence of a discrete, $n$-fold rotational symmetry. The total crystal angular momentum is conserved (modulo an integer multiple of $n$) in scattering processes and gives rise to selection rules for the scattering of excitons with other quasi-particles. Moreover, we demonstrate how crystal symmetries can be exploited to improve the computational efficiency of calculations with the Bethe–Salpeter equation (BSE). The methodology is developed in the general case of spinorial systems and includes the case of non-hermitian BSE Hamiltonians (for situations where the Tamm-Dancoff approximation cannot be applied). The methodology is validated across three prototypical systems hosting Wannier-type, Frenkel-type, and mixed Frenkel–Wannier excitons, highlighting its generality and robustness. Overall, this framework provides a precise understanding of exciton symmetries and the resulting optical and phonon selection rules, offering a foundation for future studies on tailoring the optical properties of materials through symmetry considerations.

\section{Acknowledgements}
We acknowledge the use of the HPC facilities of the University of Luxembourg~\cite{VCPKVO_HPCCT22}. N.M and L.W acknowledge funding by the FNR through project C22/MS/17415967/ExcPhon.
F.P. acknowledges funding by ICSC - Centro Nazionale di Ricerca in High Performance Computing, Big Data and Quantum Computing – funded by the European Union through the Italian Ministry of University and Research under PNRR M4C2I1.4 (Grant No. CN00000013).
DS acknowledges funding from the ``MAterials design at the eXascale'' (MaX) center of excellence, co-funded by the European High Performance Computing joint Undertaking (JU) and participating countries (Grant Agreement No. 101093374), and from the PRIN project ``Exploring extreme ultraviolet excitons with attosecond time resolution'' (EXATTO), Grant No. 2022PX279E from MIUR (Italy).
N.M thanks Dhruv Sharma, Andrea Marini, Aseem Rajan Kshirsagar, Henry Fried and Riccardo Reho for helpful discussions. This work is dedicated to the memory of Fran\c{c}ois Ducastelle, with whom we had the pleasure of working on the symmetry of excitons in hBN~\cite{PhysRevB.94.125303}. The present study aims to systematically address the many questions about exciton symmetries that he raised over the years.

\emph{Note added.}--During the preparation of our manuscript, an  article appeared~\cite{Bajaj2025}, which assigns irreducible-representation labels for excitonic states using the Koster–Dimmock–Wheeler–Statz notation~\cite{KDWS_notation} and focuses on the block diagonalization of the Bethe–Salpeter equation Hamiltonian.

\emph{Note added in proofs.}--During the preparation of page proofs, we became aware of Ref.~\cite{Stohler2026Mar}, which demonstrates how symmetry-based block-diagonalization can be employed to accelerate the diagonalization of the Bethe--Salpeter Hamiltonian.

\section{Data Availability}
The data that support the findings of this article are openly available on NOMAD repository~\cite{nomad_2026_02_27}

\onecolumngrid
\appendix

\section{Invariance of the kernel of the Bethe–Salpeter equation}
\label{section:bse_kernel_invar}
Consider an $n$-particle Green's function~\cite{Stefanucci2013}:
\begin{equation}
G^{n}(1,2,\ldots,n,\bar{1},\bar{2},\ldots,\bar{n}) = (-i)^n \bra{0}T\{ \hat{\psi}(1)\hat{\psi}(2)\cdots\hat{\psi}(n)\hat{\psi}^\dagger(\bar{1})\hat{\psi}^\dagger(\bar{2})\cdots\hat{\psi}^\dagger(\bar{n}) \} \ket{0},
\label{eq:Green_n}
\end{equation}
where $T$ is the time-ordering operator, $\ket{0}$ is the ground state, $\hat{\psi}(1)$ is the electron field operator, and the numbers denote position, spin, and time coordinates, e.g., $1 \to (\mathbf{r}_1\sigma_1,t_1)$. The ground state is invariant under all crystal symmetries, i.e., $\hat{U}(g)\ket{0} = \hat{U}^\dagger(g)\ket{0} = \ket{0}$. This implies
\begin{equation}
\begin{aligned}
G^{n}(1,2,\ldots,n,\bar{1},\bar{2},\ldots,\bar{n}) &= (-i)^n \bra{0} \hat{U}(g) T\{ \hat{\psi}(1)\hat{\psi}(2)\cdots\hat{\psi}(n)\hat{\psi}^\dagger(\bar{1})\hat{\psi}^\dagger(\bar{2})\cdots\hat{\psi}^\dagger(\bar{n}) \} \hat{U}^\dagger(g)\ket{0} \\
&= (-i)^n \bra{0} T\{ \hat{U}(g)\hat{\psi}(1)\hat{U}^\dagger(g) \, \hat{U}(g)\hat{\psi}(2)\hat{U}^\dagger(g) \cdots \hat{U}(g)\hat{\psi}(n)\hat{U}^\dagger(g) \\
&\quad\quad\quad\quad\quad\quad \hat{U}(g)\hat{\psi}^\dagger(\bar{1})\hat{U}^\dagger(g) \, \hat{U}(g)\hat{\psi}^\dagger(\bar{2})\hat{U}^\dagger(g) \cdots \hat{U}(g)\hat{\psi}^\dagger(\bar{n})\hat{U}^\dagger(g) \} \ket{0} \\
&= (-i)^n \bra{0} T\{ \hat{\psi}(1')\hat{\psi}(2')\cdots\hat{\psi}(n')\hat{\psi}^\dagger(\bar{1}')\hat{\psi}^\dagger(\bar{2}')\cdots\hat{\psi}^\dagger(\bar{n}') \} \ket{0} \\
&= G^{n}(1',2',\ldots,n',\bar{1}',\bar{2}',\ldots,\bar{n}'),
\end{aligned} 
\label{eq:Green_n_rot1}
\end{equation}
where we used $\hat{U}(g)\hat{\psi}(1)\hat{U}^\dagger(g) = \hat{\psi}(1')$, with $1' \to (\mathbf{r}_1'\sigma_1',t_1)$ and $\mathbf{r}_1' = R^{-1}(\mathbf{r}_1-\bm{\tau})$. 

In the case of time-reversal symmetry, Eq.~\eqref{eq:Green_n_rot1} becomes
\begin{equation}
G^{n}(1,2,\ldots,n,\bar{1},\bar{2},\ldots,\bar{n})
    = (\bar{G}^{n}(1',2',\ldots,n',\bar{1}',\bar{2}',\ldots,\bar{n}'))^*,
\label{eq:Green_n_rot2}
\end{equation}
with $1' \to (\mathbf{r}_1\sigma_1',-t_1)$ and
$\bar{G}$ is the anti-time-ordered Green’s function. Here we used $\hat{U}(\mathscr{T})\hat{\psi}(1)\hat{U}^\dagger(\mathscr{T}) = \hat{\psi}(1')$ 
and the conjugation arises due to the anti-unitary property of the time-reversal symmetry~\cite{Kitaev2009May}.

Equations~\eqref{eq:Green_n_rot1} and \eqref{eq:Green_n_rot2} show that the Green's functions are invariant under all crystal symmetry operations and time-reversal symmetry~(if the latter applies). This implies that the two-particle correlation functions, defined as~\cite{Martin_Reining_Ceperley_2016} 
\begin{equation}
\begin{aligned}
    &L(1, 2; 3, 4) = G^{(2)}(1, 2; 3, 4) + G(1, 2) G(4, 3),\\
    &L_0(1, 2; 3, 4) = G(1,3)G(4,2),
\end{aligned}
    \label{eq:2part_correlation_function}
\end{equation}
are also invariant under all crystal symmetry operations. 
Now consider the BSE as given in Eq.~\eqref{eq:bse}:
\begin{equation}
L(1,2;3,4) = L_0(1,2;3,4) + \int d(5,6,7,8)\, L_0(1,2;5,6)K(5,6;7,8)L(7,8;3,4).
\label{eq:bse_appendix}
\end{equation}
We write the BSE for rotated coordinates by replacing every coordinate $n$ by $n'$ in Eq.~\eqref{eq:bse_appendix}, which gives
\begin{equation}
L(1',2';3',4') = L_0(1',2';3',4') + \int d(5',6',7',8')\, L_0(1',2';5',6')K(5',6';7',8')L(7',8';3',4').
\label{eq:bse_appendix_rotated}
\end{equation}

Since the volume element is invariant under crystal symmetry operations, we have $d(5',6',7',8') = d(5,6,7,8)$.  
Substituting this into Eq.~\eqref{eq:bse_appendix_rotated}, we obtain
\begin{equation}
L(1',2';3',4') = L_0(1',2';3',4') + \int d(5,6,7,8)\, L_0(1',2';5',6')K(5',6';7',8')L(7',8';3',4').
\label{eq:bse_appendix_rotated1}
\end{equation}

Since $L$ and $L_0$ are invariant under all crystal symmetry operations, i.e.,
\begin{align*}
L(1',2';3',4') &= L(1,2;3,4), \\
L_0(1',2';3',4') &= L_0(1,2;3,4), \\
L_0(1',2';5',6') &= L_0(1,2;5,6), \\
L(7',8';3',4') &= L(7,8;3,4).
\end{align*}

Substituting these relations into Eq.~\eqref{eq:bse_appendix_rotated1}, we obtain
\begin{equation}
L(1,2;3,4) = L_0(1,2;3,4) + \int d(5,6,7,8)\, L_0(1,2;5,6)K(5',6';7',8')L(7,8;3,4).
\label{eq:bse_appendix_rotated2}
\end{equation}

We now have two expressions for $L(1,2;3,4)$: the original Eq.~\eqref{eq:bse_appendix} and the transformed Eq.~\eqref{eq:bse_appendix_rotated2}. Since they must be equal, we can equate their right-hand sides i.e.,
\begin{equation}
\int d(5,6,7,8)\, L_0(1,2;5,6)K(5,6;7,8)L(7,8;3,4) 
= \int d(5,6,7,8)\, L_0(1,2;5,6)K(5',6';7',8')L(7,8;3,4).
\label{eq:bse_rhs_comp_rotated}
\end{equation}

After rearranging Eq.~\eqref{eq:bse_rhs_comp_rotated}, we obtain
\begin{equation}
\int d(5,6,7,8)\, L_0(1,2;5,6)\big[ K(5,6;7,8) - K(5',6';7',8') \big] L(7,8;3,4) = 0.
\end{equation}

For this integral to vanish for arbitrary non-trivial functions $L_0$ and $L$, the term in square brackets must vanish i.e.,
$K(5,6;7,8) - K(5',6';7',8') = 0$. This implies 
\begin{equation}
    K(5,6;7,8) = K(5',6';7',8')
\end{equation}
and shows the invariance of the kernel $K$ under all crystal symmetry operations.

\section{Unitary properties of $\mathcal{U}(g)$}
\label{section:Uisunitary}
In this Appendix, we show that the $\mathcal{U}(g)$ matrices defined in Eq.~\eqref{eq:exe_rep_mat} are unitary, where $g$ is an element of space group $\mathcal{G}$. Let $n = \{ \mathbf{k_1}, \mathbf{k_2}, a,b\}$ and consider $\mathcal{U}(g) \mathcal{U}(g)^\dagger$:

\begin{equation}
    \begin{aligned}
    \mathcal{U}(g) \mathcal{U}(g)^\dagger &= \quad \sum _n \mathcal{U}^{\vphantom{*}}_{n',n}(g) \mathcal{U}^*_{n'',n}(g) \\ 
    &= \ \sum_{\mathbf{k_1}, \mathbf{k_2}, a,b} 
    \mathcal{D}^{\vphantom{*}}_{\mathbf{k}_1,a'a}(g) \mathcal{D}^*_{\mathbf{k}_2,b'b}(g) 
    \delta^{\vphantom{*}}_{R\mathbf{k}^{\vphantom{*}}_1,\mathbf{k}_1'} \delta^{\vphantom{*}}_{R\mathbf{k}^{\vphantom{*}}_2,\mathbf{k}_2'} \mathcal{D}^*_{\mathbf{k}^{\vphantom{*}}_1,a''a}(g) \mathcal{D}^{\vphantom{*}}_{\mathbf{k}^{\vphantom{*}}_2,b''b}(g) 
    \delta^{\vphantom{*}}_{R\mathbf{k}^{\vphantom{*}}_1,\mathbf{k}_1''} \delta^{\vphantom{*}}_{R\mathbf{k}^{\vphantom{*}}_2,\mathbf{k}_2''} \\ 
    &= \quad \delta^{\vphantom{*}}_{\mathbf{k}_2'', \mathbf{k}_2'} \delta^{\vphantom{*}}_{\mathbf{k}_1'', \mathbf{k}_1'} 
    \sum_{ a,b} \mathcal{D}^{\vphantom{*}}_{R^{-1}\mathbf{k}_1',a'a}(g) \mathcal{D}^*_{R^{-1}\mathbf{k}_2',b'b}(g) \mathcal{D}^*_{R^{-1}\mathbf{k}_1',a''a}(g) \mathcal{D}^{\vphantom{*}}_{R^{-1}\mathbf{k}_2',b''b}(g) \\ 
    &= \quad \delta_{\mathbf{k}_2'', \mathbf{k}_2'} \delta_{\mathbf{k}_1'', \mathbf{k}_1'} 
    \delta_{a'', a'} \delta_{b'', b'} \\
    &= \quad \delta_{n',n''}.
    \label{eq:Umat_uni}
    \end{aligned}
\end{equation}

In the above equation, we used the unitary property of the $\mathcal{D}(g)$ matrices.

\section{Proof that the set of $\mathcal{U}(g)$ forms a linear representation of the Space Group.}
\label{section:U_is_rep_of_G}
In this Appendix, we show that the set of all $\mathcal{U}(g)$ matrices  $\forall \ g \in \mathcal{G}$ , where $\mathcal{G}$ is the space group defined in Eq.~\eqref{eq:exe_rep_mat}, forms a representation of the space group. Let $g_1 = \{ R_1 \ | \ \boldsymbol{\tau}_1\}$ and $g_2 = \{ R_2 \ | \ \boldsymbol{\tau}_2\}$ be two elements in $\mathcal{G}$. The product $g_1\cdot g_2 = \{ R_1R_2 \ | \ R_1\boldsymbol{\tau}_2 + \boldsymbol{\tau}_1\}$.
In order to show that the $\mathcal{U}(g)$ matrices form a linear representation of the space group, we need to show that 
\begin{equation}
    \mathcal{U}(g_1)\mathcal{U}(g_1) = \mathcal{U}(g_1\cdot g_2) \ \forall g_1,g_2 \in \mathcal{G}
    \label{eq:Urep_proof1}
\end{equation}
To prove Eq.~\eqref{eq:Urep_proof1}, we consider $\mathcal{U}(g_1)\mathcal{U}(g_2)$ and let $n = \{ \mathbf{k_1}, \mathbf{k_2}, a,b\}$,
\begin{equation}
    \begin{aligned}
        (\mathcal{U}(g_1)\mathcal{U}(g_2))_{n',n''} & = \quad \sum_n\mathcal{U}_{n',n}(g_1)\mathcal{U}_{n,n''}(g_2)\\
        &= \ \sum_{\mathbf{k_1}, \mathbf{k_2}, a,b} \Big\{ \mathcal{D}^{\vphantom{*}}_{\mathbf{k}^{\vphantom{*}}_1,a'a}(g_1) \mathcal{D}^*_{\mathbf{k}^{\vphantom{*}}_2,b'b}(g_1)  \delta^{\vphantom{*}}_{R^{\vphantom{*}}_1\mathbf{k}^{\vphantom{*}}_1,\mathbf{k}_1'} \delta^{\vphantom{*}}_{R^{\vphantom{*}}_1\mathbf{k}^{\vphantom{*}}_2,\mathbf{k}_2'}    \mathcal{D}^{\vphantom{*}}_{\mathbf{k}''_1,a^{\vphantom{*}}a''}(g_2) \mathcal{D}^*_{\mathbf{k}''_2,b^{\vphantom{*}}b''}(g_2) \delta^{\vphantom{*}}_{R^{\vphantom{*}}_2\mathbf{k}''_1,\mathbf{k}^{\vphantom{*}}_1} \delta^{\vphantom{*}}_{R^{\vphantom{*}}_2\mathbf{k}''_2,\mathbf{k}^{\vphantom{*}}_2} 
        \Big \} \\ &
        = \quad \sum_{a,b} \Big\{ \mathcal{D}^{\vphantom{*}}_{R^{\vphantom{*}}_2\mathbf{k}''_1,a'a}(g_1) \mathcal{D}^*_{R^{\vphantom{*}}_2\mathbf{k}''_2,b'b}(g_1)  \mathcal{D}^{\vphantom{*}}_{\mathbf{k}''_1,a^{\vphantom{*}}a''}(g_2) 
         \mathcal{D}^*_{\mathbf{k}''_2,b^{\vphantom{*}}b''}(g_2)
        \delta^{\vphantom{*}}_{R^{\vphantom{*}}_1R^{\vphantom{*}}_2\mathbf{k}''_1,\mathbf{k}'_1} \delta^{\vphantom{*}}_{R^{\vphantom{*}}_1R^{\vphantom{*}}_2\mathbf{k}''_2,\mathbf{k}'_2} 
        \Big \}
        \label{eq:Urep_proof2}
    \end{aligned}
\end{equation}

Now, consider the following product for the $\mathcal{D}$ matrices as given in Eq.~\eqref{eq:Urep_proof2}
\begin{equation}
    \begin{aligned}
    \sum_{a} \mathcal{D}_{R^{\vphantom{*}}_2 \mathbf{k}''_1, a'a} (g^{\vphantom{*}}_1) \mathcal{D}_{\mathbf{k}''_1, a^{\vphantom{*}}a''}(g^{\vphantom{*}}_2) 
    &= \sum_{a} \langle R^{\vphantom{*}}_1 R^{\vphantom{*}}_2 \mathbf{k}''_1,  a' | \hat{U}(g^{\vphantom{*}}_1)| R^{\vphantom{*}}_2 \mathbf{k}''_1,  a \rangle  \langle R^{\vphantom{*}}_2 \mathbf{k}''_1,  a | \hat{U}(g^{\vphantom{*}}_2)| \mathbf{k}''_1,  a'' \rangle
     \\&
    = \langle R^{\vphantom{*}}_1 R^{\vphantom{*}}_2 \mathbf{k}''_1,  a' | \hat{U}(g^{\vphantom{*}}_1) \hat{U}(g^{\vphantom{*}}_2)| \mathbf{k}''_1,  a'' \rangle
      \\&
    =\langle R^{\vphantom{*}}_1 R^{\vphantom{*}}_2 \mathbf{k}''_1,  a' | \hat{U}(g^{\vphantom{*}}_1 \cdot g^{\vphantom{*}}_2) e^{i\phi(g_1,g_2)}| \mathbf{k}''_1,  a'' \rangle \\&
    = e^{i\phi(g_1,g_2)}  \mathcal{D}_{\mathbf{k}''_1, a'a''} (g^{\vphantom{*}}_1\cdot g^{\vphantom{*}}_2)
    \label{eq:Dmat_gprod}
    \end{aligned}
\end{equation}
where it is assumed that the set of \( \hat{U}(g) \) operators furnish a projective representation of the space group, i.e., $\hat{U}(g_1) \hat{U}(g_2) = e^{i\phi(g_1,g_2)} \hat{U}(g_1 \cdot g_2)$. Substituting Eq.~\eqref{eq:Dmat_gprod} in Eq.~\eqref{eq:Urep_proof2}, we obtain
\begin{equation}
\begin{aligned}
    \left(\mathcal{U}(g_1)\mathcal{U}(g_2)\right)^{\vphantom{*}}_{n',n''} &= 
    \delta^{\vphantom{*}}_{R^{\vphantom{*}}_1R^{\vphantom{*}}_2\mathbf{k}''_1,\mathbf{k}'_1}
        \delta^{\vphantom{*}}_{R^{\vphantom{*}}_1R^{\vphantom{*}}_2\mathbf{k}''_2,\mathbf{k}'_2} 
    e^{i\phi(g_1,g_2)} e^{-i\phi(g_1,g_2)} 
      \mathcal{D}^{\vphantom{*}}_{\mathbf{k}''_1, a'a''} (g_1\cdot g_2)  \mathcal{D}^*_{\mathbf{k}''_2, b'b''} (g_1\cdot g_2) \\ &
       = \mathcal{U}_{n',n''}(g_1 \cdot g_2)
    \label{eq:Urep_proof3}
\end{aligned}
\end{equation}
This implies that the set of $\mathcal{U}(g) \ \forall g \in \mathcal{G}$ form a linear representation of the space group $\mathcal{G}$.

\section{Exciton phonon matrix elements}\label{app:exc-ph_definition}
\label{appen:ex-ph_def}
We consider the following matrix element within the Tamm-Dancoff approximation~\cite{tda_ref}:
\begin{equation}
    \tilde{\mathcal{G}}^{\lambda}_{\vphantom{\mathbf{q}}S',S}(\mathbf{Q},\mathbf{q}) \equiv \langle S', \mathbf{q}\!+\! \mathbf{Q} \ | \partial^\lambda_\mathbf{q} V | \ S, \mathbf{Q} \rangle  
    \label{eq:exph_braket}
\end{equation}
where $\mathbf{Q}$ is the transfer momentum of the initial exciton $S$, $\mathbf{q}$ is the crystal momentum of the phonon with mode index $\lambda$, and $\mathbf{Q} \!+\!\mathbf{q}$ is the transfer momentum of the outgoing exciton $S'$. \replytoreferee{The operator $\partial^{\lambda}_{\mathbf{q}} V$ denotes the directional derivative of the self-consistent potential along the displacement vector (with dimensions of length) of the phonon mode with index $\lambda$. It is defined as
\begin{equation}
    \partial^{\lambda}_{\mathbf{q}} V =
    \sqrt{\frac{\hbar}{2\omega^{\lambda}}}
    \sum_{\kappa\alpha}
    \frac{\tilde{e}^{\lambda}_{\mathbf{q},\kappa\alpha}}{\sqrt{M_{\kappa}}}
    \frac{\partial V}{\partial R_{\kappa\alpha}},
    \label{eq:dvq_def}
\end{equation}
where $M_{\kappa}$ is the mass of atom $\kappa$ located at $\mathbf{R}_{\kappa}$, $\tilde{e}^{\lambda}_{\mathbf{q},\kappa\alpha}$ is the $\lambda$-th eigenvector of the phonon dynamical matrix at crystal momentum $\mathbf{q}$, $\alpha$ denotes the cartesian direction, and $\omega^{\lambda}$ is the frequency of the $\lambda$-th phonon mode.
}

The excitonic state $\ket{S,\mathbf{Q}}$ can be written as~\cite{PhysRevLett.81.2312,PhysRevB.62.4927,PhysRevLett.80.4510}:
\begin{equation}
| S,\mathbf{Q} \rangle = \sum_{\mathbf{k}cv} X^{S;\mathbf{Q}\vphantom{\dagger}}_{\mathbf{k}cv} a^\dagger_{\mathbf{k}c}a^{\vphantom{\dagger}}_{\mathbf{k}-\mathbf{Q}v}|0\rangle,
\label{eq:exe_state}    
\end{equation}
where $\ket{0}$ is the non-interacting ground state.

Moreover, the \replytoreferee{operator $\partial^{\lambda}_{\mathbf{q}} V$ is} expressed in terms of one-particle Bloch states, which is given by
\begin{equation}
\begin{aligned}
    \partial^\lambda_{\vphantom{\tilde{\mathbf{k}}}\mathbf{q}} V &= \sum_{m,n,\tilde{\mathbf{k}}}\langle m, \mathbf{q}\!+\! \tilde{\mathbf{k}} \ | \partial^\lambda_{\vphantom{\tilde{\mathbf{k}}}\mathbf{q}} V | \ n, \tilde{\mathbf{k}} \rangle \ a^\dagger_{\vphantom{\tilde{\mathbf{k}}}\mathbf{k}+\mathbf{q}m}a^{\vphantom{\dagger}}_{\vphantom{\tilde{\mathbf{k}}}\mathbf{k}n} 
    = \sum_{m,n,\tilde{\mathbf{k}}} a^\dagger_{\tilde{\mathbf{k}}+\mathbf{q}m}a^{\vphantom{\dagger}}_{\tilde{\mathbf{k}}n} \tilde{g}^{\lambda}_{\vphantom{\tilde{\mathbf{k}}}m,n}(\tilde{\mathbf{k}},\mathbf{q}),
\end{aligned}
    \label{eq:PHdeformationKS}
\end{equation}
where $\tilde{g}^{\lambda}_{m,n}(\tilde{\mathbf{k}},\mathbf{q}) =  \langle m, \mathbf{q}+\tilde{\mathbf{k}} \mid
    \partial^\lambda_{\mathbf{q}} V
    \mid n, \tilde{\mathbf{k}} \rangle$ are the electron-phonon matrix elements.

Substituting Eq.~\eqref{eq:exe_state} and Eq.~\eqref{eq:PHdeformationKS} into Eq.~\eqref{eq:exph_braket} gives
\begin{equation}
\begin{aligned}
    \tilde{\mathcal{G}}^{\lambda}_{S',S}(\mathbf{Q},\mathbf{q}) =\sum_{\mathbf{k}\mathbf{k^\prime} \mathbf{\tilde{k}} cc^\prime vv^\prime mn} \left\{ \left(X^{\vphantom{\dagger}S';(\mathbf{Q} + \mathbf{q}) }_{\vphantom{\mathbf{\tilde{k}}}\mathbf{k^\prime}c^\prime v^\prime} \right)^* X^{\vphantom{\dagger}S;(\mathbf{Q})}_{\vphantom{\mathbf{\tilde{k}}}\mathbf{k}cv} \tilde{g}^{\lambda}_{m,n}(\tilde{\mathbf{k}},\mathbf{q}) \  \langle 0 |a^\dagger_{\vphantom{\mathbf{\tilde{k}}}\mathbf{k^\prime}-\mathbf{Q}-\mathbf{q}v^\prime} a^{\vphantom{\dagger}}_{\vphantom{\mathbf{\tilde{k}}}\mathbf{k^\prime}c^\prime} a^\dagger_{\mathbf{\tilde{k} + \mathbf{q}}m} a^{\vphantom{\dagger}}_{\mathbf{\tilde{k}}n} a^\dagger_{\vphantom{\mathbf{\tilde{k}}}\mathbf{k}c} a^{\vphantom{\dagger}}_{\vphantom{\mathbf{\tilde{k}}}\mathbf{k}-\mathbf{Q}v}|0\rangle
 \right\} .
\end{aligned}
\label{eq:exph_expansion}
\end{equation}

Consider the correlation function in Eq.~\eqref{eq:exph_expansion}, and employing Wick's contractions gives
\begin{equation}
\begin{aligned}
    &\langle 0 |a^\dagger_{\vphantom{\mathbf{\tilde{k}}}\mathbf{k^\prime}-\mathbf{Q}-\mathbf{q}v^\prime} a^{\vphantom{\dagger}}_{\vphantom{\mathbf{\tilde{k}}}\mathbf{k^\prime}c^\prime} a^\dagger_{\mathbf{\tilde{k} + \mathbf{q}}m} a^{\vphantom{\dagger}}_{\mathbf{\tilde{k}}n} a^\dagger_{\vphantom{\mathbf{\tilde{k}}}\mathbf{k}c} a^{\vphantom{\dagger}}_{\vphantom{\mathbf{\tilde{k}}}\mathbf{k}-\mathbf{Q}v}|0\rangle \\=& \left\{ 
    -\delta_{\mathbf{k}-\mathbf{Q}, \mathbf{\tilde{k} + \mathbf{q}} \vphantom{\tilde{k}, k, v, c, m, n}}
    \delta_{ v ,m \vphantom{\tilde{k}, k, v, c, m, n}}
    \delta_{\mathbf{k^\prime},\mathbf{k} \vphantom{\tilde{k}, k, v, c, m, n}}
    \delta_{c^\prime, c \vphantom{\tilde{k}, k, v, c, m, n}}
    \delta_{\mathbf{\tilde{k}},\mathbf{k^\prime}-\mathbf{Q}-\mathbf{q} \vphantom{\tilde{k}, k, v, c, m, n}}
    \delta_{ v',n \vphantom{\tilde{k}, k, v, c, m, n}}  \right.
    \\&+\delta_{\mathbf{k^\prime}-\mathbf{Q}-\mathbf{q},\mathbf{k}-\mathbf{Q} \vphantom{\tilde{k}, k, v, c, m, n}}
    \delta_{ v^\prime, v \vphantom{\tilde{k}, k, v, c, m, n}}
    \delta_{\mathbf{k^\prime}, \mathbf{\tilde{k}} + \mathbf{q}  \vphantom{\tilde{k}, k, v, c, m, n}}
    \delta_{c^\prime, m \vphantom{\tilde{k}, k, v, c, m, n}}
    \delta_{\mathbf{\tilde{k}},\mathbf{k} \vphantom{\tilde{k}, k, v, c, m, n}}
    \delta_{ n,c  \vphantom{\tilde{k}, k, v, c, m, n}}
    \\& \left.
    +\delta_{\mathbf{k^\prime}-\mathbf{Q}-\mathbf{q} ,\mathbf{k}-\mathbf{Q}  \vphantom{\tilde{k}, k, v, c, m, n}} 
    \delta_{ v^\prime, v  \vphantom{\tilde{k}, k, v, c, m, n}}
    \delta_{\mathbf{k^\prime}, \mathbf{k}  \vphantom{\tilde{k}, k, v, c, m, n}}
    \delta_{c^\prime, c  \vphantom{\tilde{k}, k, v, c, m, n}}
    \delta_{\mathbf{\tilde{k} + \mathbf{q}}, \mathbf{\tilde{k}} \vphantom{\tilde{k}, k, v, c, m, n}}
    \delta_{m,n  \vphantom{\tilde{k}, k, v, c, m, n}}\right\}.
\end{aligned}
\label{eq:4pt_expandedQ}
\end{equation}

Substituting Eq.~\eqref{eq:4pt_expandedQ} into Eq.~\eqref{eq:exph_expansion}, we obtain
\begin{equation}
    \tilde{\mathcal{G}}^{\lambda}_{S',S}(\mathbf{Q},\mathbf{q}) = \mathcal{G}^{\lambda}_{S',S}(\mathbf{Q},\mathbf{q}) + \delta^{\vphantom{\lambda}}_{\mathbf{q},0} \delta^{\vphantom{\lambda}}_{S,S'} \sum_{m,\mathbf{k}} \tilde{g}^\lambda_{m,m}(\mathbf{k},\mathbf{q}\!=\!0),
    \label{eq:exph_prefinal}
\end{equation}
where $\mathcal{G}^{\lambda}_{S',S}(\mathbf{Q},\mathbf{q})$ are the exciton-phonon matrix elements. The extra term in Eq.~\eqref{eq:exph_prefinal} corresponds to a disconnected diagram and is canceled when performing a perturbation expansion due to normalization (For example, see the Supplementary Information of Ref.~\cite{doi:10.1126/sciadv.abb5915} for the case of resonant Raman matrix elements; see also Refs.~\cite{Paleari2019,PhysRevB.102.045136,PhysRevResearch.2.012032}.).
From Eq.~\eqref{eq:4pt_expandedQ} and Eq.~\eqref{eq:exph_expansion}, the exciton-phonon matrix elements $\mathcal{G}^{\lambda}_{S',S}(\mathbf{Q},\mathbf{q})$ are given by~\cite{doi:10.1126/sciadv.abb5915}
\begin{equation}
\begin{aligned}
    \mathcal{G}^{\vphantom{\dagger}\lambda}_{S',S}(\mathbf{Q},\mathbf{q}) =& 
     \sum_{\mathbf{k} cc^\prime v}  \left(X^{\vphantom{\dagger}S';(\mathbf{Q} + \mathbf{q}) }_{\mathbf{k}+\mathbf{q}c^\prime v} \right)^* X^{\vphantom{\dagger}S;(\mathbf{Q})}_{\mathbf{k}cv} \tilde{g}^{\vphantom{\dagger}\lambda}_{c',c}(\mathbf{k},\mathbf{q}) 
  -
 \sum_{\mathbf{k}  c vv^\prime}  \left(X^{\vphantom{\dagger}S';(\mathbf{Q} + \mathbf{q}) }_{\mathbf{k}c v^\prime} \right)^* X^{\vphantom{\dagger}S;(\mathbf{Q})}_{\mathbf{k}cv} \tilde{g}^{\vphantom{\dagger}\lambda}_{v,v'}(\mathbf{k} \!-\!\mathbf{Q}\!-\!\mathbf{q},\mathbf{q}) .
 \end{aligned}
    \label{eq:exph_matr_elmentQ}
\end{equation}

Note that the specific internal $\mathbf{Q}$- and $\mathbf{q}$-momentum dependence of the above expression depends on the chosen conventions for momentum conservation between initial and final states. Here, the electronic transitions have momentum transfer $\mathbf{k}\!-\!\mathbf{Q} \rightarrow \mathbf{k}$ while the electron-phonon scattering has $\mathbf{k}\rightarrow \mathbf{k}\!+\!\mathbf{q}$. If the latter were defined with the convention $\mathbf{k}\!-\!\mathbf{q}\rightarrow \mathbf{k}$, then the form of the Eq.~\eqref{eq:exph_matr_elmentQ} would change. Different many-body codes may provide by default excitonic and electron-phonon input data following different momentum flow conventions than those chosen here.

\section{Rotation of electron-phonon matrix elements}
\label{app:rotate_elph_me}
In this section, we demonstrate how electron-phonon matrix elements transform under symmetry operations. The action of the symmetry operator $\hat{U}(g)$ on \replytoreferee{ $\partial^{\lambda}_{\mathbf{q}} V(\mathbf{r})$ (represented in position basis)} is given by~\cite{PhysRevB.76.165108,RevModPhys.40.1}

\begin{equation}
    \hat{U}(g)\partial_{\mathbf{q}}^\lambda V_{\text{scf}}(\mathbf{r})\hat{U}^\dagger(g) = \sum_{\lambda'}\Gamma^{\vphantom{\lambda'}}_{\mathbf{q},\lambda'\lambda}(g)\partial_{R\mathbf{q}}^{\lambda'} V_{\text{scf}}(\mathbf{r}),
    \label{eq:deformationPot}
\end{equation}

Where $g$ represents a spatial crystal symmetry operation or time-reversal symmetry. If $g$ is a spatial crystal symmetry operation, it corresponds to a coordinate transformation $\mathbf{r} \rightarrow R\mathbf{r} + \mathbf{v}$, with $R$ being an orthogonal matrix and $\mathbf{v}$ a translation vector.

The term $\Gamma_{\mathbf{q},\lambda'\lambda}(g)$ is the phase matrix (or representation matrix, if $g$ belongs to the little group of $\mathbf{q}$) for the phonon modes, analogous to the $\mathcal{D}$ matrices for Bloch states. It is unitary and block diagonal in degenerate subspaces when phonon eigenvectors are chosen to be orthogonal. When $g$ is a spatial symmetry, $\Gamma_{\mathbf{q},\lambda'\lambda}(g)$ is given by~\cite{RevModPhys.40.1}

\begin{equation}
\begin{aligned}
    \Gamma^{\vphantom{*}}_{\mathbf{q},\lambda'\lambda}(g) &= 
    \left(\mathbf{d}_{R\mathbf{q}}^{\lambda'}\right)^\dagger \hat{U}(g) \mathbf{d}^\lambda_{\mathbf{q}} 
    = \sum_{\kappa,\beta,\tilde{\kappa},\alpha} e^{i \mathbf{q} \cdot (g^{-1} \boldsymbol{\tau}_{\tilde{\kappa}} - \boldsymbol{\tau}_\kappa)} R^{\vphantom{*}}_{\alpha \beta} d^\lambda_{\mathbf{q},\kappa,\beta} \left(d^{\lambda'}_{R\mathbf{q},\tilde{\kappa},\alpha}\right)^*.
\end{aligned}
    \label{eq:ph_rep}
\end{equation}

Here, $\mathbf{d}_{R\mathbf{q}}^{\lambda'}$ and $\mathbf{d}^\lambda_{\mathbf{q}}$ are the phonon eigenvectors for $\mathbf{q}$ and $R\mathbf{q}$ phonon crystal momenta. In the case of time-reversal symmetry, we have

\begin{equation}
\begin{aligned}
    \Gamma^{\vphantom{*}}_{\mathbf{q},\lambda'\lambda}(g) &= 
    (\mathbf{d}_{-\mathbf{q}}^{\lambda'})^\dagger (\mathbf{d}^\lambda_{\mathbf{q}})^*.
\end{aligned}
    \label{eq:ph_rep_trev}
\end{equation}

If $R\mathbf{q} \neq \mathbf{q} + \mathbf{G}$, where $\mathbf{G}$ is a reciprocal lattice vector, and the phonon eigenvector at $R\mathbf{q}$ is obtained by applying the symmetry operation $g$ to the eigenvector at $\mathbf{q}$, then $\Gamma_{\mathbf{q}}(g)$ is an identity matrix.

It is important to note that the \replytoreferee{operator $\partial^{\lambda}_{\mathbf{q}} V$} is generally a $2 \times 2$ matrix in the spinor subspace. Therefore, the symmetry operators $\hat{U}(g)$ must include spin rotation matrices that account for transformations in the spinor subspace.

Now, consider the following electron-phonon matrix elements:

\begin{equation}
\begin{aligned}
     \tilde{g}^{\lambda'}_{m,n}(R\mathbf{k},R\mathbf{q}) =& \langle m, R\mathbf{q}\!+\!R\mathbf{k} \ | \left( \partial^{\lambda'}_{R\mathbf{q}} V \ | \ n, R\mathbf{k} \rangle \right),
\end{aligned}
\label{eq:elph_def001}
\end{equation}

where we use parentheses to distinguish the action of the operator on either the bra or the ket, also taking time-reversal symmetry into account.

From the definition of the phase matrices in Eq~\eqref{eq:intro_dmat}, and using their unitary property, we have:

\begin{equation}
\begin{aligned}
    &\ket{n,R\mathbf{k}} = \sum_{n'} \mathcal{D}_{\mathbf{k},nn'}^*(g) \hat{U}(g) \ket{n',\mathbf{k}}, \\
    &\ket{m,R\mathbf{k}\!+\!R\mathbf{q}} = \sum_{m'} \mathcal{D}_{\mathbf{k}+\mathbf{q},mm'}^*(g) \hat{U}(g) \ket{m',\mathbf{k} \!+\! \mathbf{q}}.
\end{aligned}
\label{eq:phase_mat_defelph}
\end{equation}

Substituting Eq.~\eqref{eq:phase_mat_defelph} into Eq.~\eqref{eq:elph_def001}, and using Eq.~\eqref{eq:deformationPot}, we obtain:

\begin{equation}
    \tilde{g}^{\lambda'}_{m,n}(R\mathbf{k},R\mathbf{q}) 
    = \sum_{m',n'} \Bigg\{ \mathcal{D}^{\vphantom{*}}_{\mathbf{k}+\mathbf{q},mm'}(g)
    \mathcal{D}_{\mathbf{k},nn'}^*(g) 
      \left( \langle m', \mathbf{q}\!+\!\mathbf{k} \ |\hat{U}^\dagger(g) \right)\left( \Gamma^*_{\mathbf{q},\lambda'\lambda}(g) \hat{U}(g)\partial_{\mathbf{q}}^\lambda V_{\text{scf}}(\mathbf{r})\hat{U}^\dagger(g) \hat{U}(g) | \ n', \mathbf{k} \rangle \right) \Bigg\}
\label{eq:elph_rot111}
\end{equation}

If $g$ is a normal spatial symmetry, we have:

\begin{equation}
\begin{aligned}
    \tilde{g}^{\lambda'}_{m,n}(R\mathbf{k},R\mathbf{q}) = &  \sum_{m',n',\lambda'} \Gamma^*_{\mathbf{q},\lambda'\lambda}(g)\mathcal{D}^{\vphantom{*}}_{\mathbf{k}+\mathbf{q},mm'}(g) 
    \mathcal{D}_{\mathbf{k},nn'}^*(g) \tilde{g}^{\lambda}_{m',n'}(\mathbf{k},\mathbf{q}) .
\end{aligned}
\label{eq:elph_rot222}
\end{equation}

In the case where $g$ is time-reversal symmetry which is anti-unitary, we need to conjugate $\tilde{g}^{\lambda}_{m',n'}(\mathbf{k},\mathbf{q})$ due to the transfer of the action of the leftmost $\hat{U}^\dagger(g)$ from the bra to the ket in Eq.~\eqref{eq:elph_rot111}, i.e.,

\begin{equation}
\begin{aligned}
    \tilde{g}^{\lambda'}_{m,n}(-\mathbf{k},-\mathbf{q}) = &  \sum_{m',n',\lambda'} \Gamma^{\vphantom{*}}_{\mathbf{q},\lambda'\lambda}(g)\mathcal{D}^{\vphantom{*}}_{\mathbf{k}+\mathbf{q},mm'}(g)  \mathcal{D}_{\mathbf{k},nn'}^*(g) (\tilde{g}^{\lambda}_{m',n'}\left(\mathbf{k},\mathbf{q})\right)^* .
\end{aligned}
\label{eq:elph_rot333}
\end{equation}

From Eqs.~\eqref{eq:elph_rot333} and \eqref{eq:elph_rot222}, we can obtain the electron-phonon matrix elements for the $R\mathbf{q}$ phonon wavevectors without explicitly evaluating the bracket. Furthermore, when $g$ belongs to the little group of $\mathbf{q}$, we can also retrieve the $R\mathbf{k}$ matrix elements from the $\mathbf{k}$ matrix elements with the correct gauge consistency.

\section{Rotation of exciton-phonon matrix elements}
\label{app:rotate_excph_me}
In this section, we show how exciton-phonon matrix elements transform within the Tamm-Dancoff approximation~\cite{tda_ref} using symmetries, in a manner similar to electron-phonon matrix elements. Consider the following bracket:
\begin{equation}
    \tilde{\mathcal{G}}^{\lambda'}_{m,n}(R\mathbf{Q},R\mathbf{q}) = \langle m, R\mathbf{q}\!+\! R\mathbf{Q} \ | \big( \partial^{\lambda'}_{R\mathbf{q}} V \ | \ n, R\mathbf{Q} \rangle \big),
    \label{eq:exph_bracket}
\end{equation}
where $\tilde{\mathcal{G}}$ represents the exciton-phonon interaction matrix elements.

Similar to the phase matrices for Bloch states, the phase matrices for excitonic states under symmetry operations are written as:
\begin{equation}
    \label{eq:phase_matExe}
    \hat{U}\ket{S,\mathbf{Q}} = \mathscr{D}_{\mathbf{Q},S'S}(g) \ket{S',R\mathbf{Q}},
\end{equation}
where \(\mathscr{D}_{\mathbf{Q},S'S}(g)\) is a unitary matrix when the excitonic states are chosen to be orthogonal. This implies that 
\begin{equation}
\begin{aligned}
    &\ket{n,R\mathbf{Q}} = \sum_{n'} \mathscr{D}_{\mathbf{Q},nn'}^*(g) \hat{U}(g) \ket{n',\mathbf{Q}}, \\
    &\ket{m,R\mathbf{Q}\!+\!R\mathbf{q}} = \sum_{m'} \mathscr{D}_{\mathbf{Q}+\mathbf{q},mm'}^*(g) \hat{U}(g) \ket{m',\mathbf{Q} \!+\! \mathbf{q}}.
\end{aligned}
\label{eq:ex_phase_mat}
\end{equation}
Following the procedure outlined for electron-phonon matrix elements in the previous section, we obtain the following transformation rules. If $g$ is a normal spatial symmetry, we have:
\begin{equation}
\begin{aligned}
    \tilde{\mathcal{G}}^{\lambda'}_{m,n}(R\mathbf{Q},R\mathbf{q}) = & \sum_{m',n',\lambda'} \Gamma^*_{\mathbf{q},\lambda'\lambda}(g) \mathscr{D}^{\vphantom{*}}_{\mathbf{Q}+\mathbf{q},mm'}(g)  \mathscr{D}_{\mathbf{Q},nn'}^*(g) \tilde{\mathcal{G}}^{\lambda}_{m',n'}(\mathbf{Q},\mathbf{q}).
\end{aligned}
\label{eq:exph_rot}
\end{equation}

If $g$ corresponds to time-reversal symmetry, we obtain:
\begin{equation}
\begin{aligned}
    \tilde{\mathcal{G}}^{\lambda'}_{m,n}(-\mathbf{Q},-\mathbf{q}) = & \sum_{m',n',\lambda'} \Gamma^{\vphantom{*}}_{\mathbf{q},\lambda'\lambda}(g) \mathscr{D}^{\vphantom{*}}_{\mathbf{Q}+\mathbf{q},mm'}(g)  \mathscr{D}_{\mathbf{Q},nn'}^*(g) (\tilde{\mathcal{G}}^{\lambda}_{m',n'}(\mathbf{Q},\mathbf{q}))^*.
\end{aligned}
\label{eq:exph_rot_TR}
\end{equation}

Substituting Eq.~\eqref{eq:exph_prefinal} into Eqs.~\eqref{eq:exph_rot_TR} and \eqref{eq:exph_rot}, we obtain an identical relation for the exciton-phonon matrix elements, where $\tilde{\mathcal{G}}$ is replaced with ${\mathcal{G}}$.

From Eqs.~\eqref{eq:exph_rot_TR} and \eqref{eq:exph_rot}, we conclude that the exciton-phonon matrix elements for $R\mathbf{q}$ phonon wavevectors can be determined without explicitly evaluating the bracket. Furthermore, when $g$ belongs to the little group of $\mathbf{q}$, we can also retrieve the $R\mathbf{Q}$ matrix elements from the $\mathbf{Q}$ matrix elements, with the correct gauge.

\section{Computational details}
\label{section:comp_details}
All ground-state DFT calculations were performed using the {\tt Quantum ESPRESSO} code~\cite{Giannozzi2017Oct}. We employed the LDA~\cite{PhysRev.136.B864} and PBE functionals~\cite{PhysRevLett.77.3865} with optimized norm-conserving Vanderbilt pseudopotentials from {\tt PseudoDojo}~\cite{PhysRevB.88.085117,vanSetten2018May} in all calculations. To obtain the equilibrium structure, we relaxed the systems with convergence thresholds of $10^{-5}$~Ry for the total energy and $10^{-5}$~Ry/Bohr for the forces. For 2D materials, to avoid spurious effects due to out-of-plane periodicity, we set the vacuum separation to at least 16~{\AA} and employed a Coulomb cutoff in all {\it ab initio} calculations~\cite{PhysRevB.96.075448}. The input parameters used for different systems in the DFT calculations are listed in Table~\ref{tab:dft_paramenters}.
\begin{table}[!ht]
    \centering
    \begin{tabular}{|c|c|c|c|}
    \hline
      System & $k$-point grid & Wavefunction cutoff (Ry) & Functional \\
      \hline
      LiF                  & $12\times12\times12$  & 120 & LDA (SR) \\
      \hline
      2D MoSe$_2$           & $12\times12\times1$ & 120 & PBE (FR) \\
      \hline
      Bulk $h$BN           & $12\times12\times4$ & 120 & LDA (SR) \\
      \hline
    \end{tabular}
    \caption{DFT input parameters for different systems. FR/SR denote full/scalar relativistic pseudopotentials. In all cases, we used $\Gamma$-centered $k$-point grids. The charge density cutoff was set to four times the wavefunction cutoff in all the cases.}
    \label{tab:dft_paramenters}
\end{table}

To obtain the phonon frequencies, eigenvectors, and the \replytoreferee{operators $\partial^{\lambda}_{\mathbf{q}} V(\mathbf{r})$}, we performed density-functional perturbation theory (DFPT) calculations using the {\tt Quantum ESPRESSO} package. We used the {\tt LetzElPhC} code~\cite{Nalabothula2025May} to compute the electron–phonon matrix elements.

Excitonic energies and wavefunctions were obtained from GW-BSE calculations using the {\tt Yambo} code~\cite{Marini2009Aug,Sangalli2019May}. For the GW step, we employed the plasmon-pole approximation~\cite{PhysRevLett.62.1169} to model the frequency-dependent dielectric function. Convergence with respect to the number of bands and $k$-points was accelerated using the {\tt G-terminator}~\cite{PhysRevB.78.085125} and, for 2D materials, the {\tt RIM-W} technique~\cite{Guandalini2023}. In all BSE calculations at $\mathbf{Q}\!=\!0$, we excluded the long-range part of the kernel, i.e., the $\mathbf{G}\!=\!0$ term of the exchange interaction~\cite{RevModPhys.74.601}. Table~\ref{tab:gw_bse_paramenters} summarizes the parameters used in the GW-BSE calculations.

\begin{table}[!ht]
    \centering
    \begin{tabular}{|c|c|c|c|c|}
    \hline
      System & $k$-point grid & Dielectric cutoff (Ry) & Bands in GW summation & Conduction/valence bands in BSE \\
      \hline
      LiF                  & $12\times12\times12$  & 30 & 80 & 1/3  \\
      \hline
      2D MoSe$_2$           & $48\times48\times1$ & 10 & 300 & 2/2 \\
      \hline
      Bulk $h$BN           & $60\times60\times4$ & 10 & 120 & 2/2 \\
      & $24\times24\times4$ (for luminescence) &  &  &\\
      \hline
    \end{tabular}
    \caption{GW-BSE input parameters for different systems. In all cases, we used $\Gamma$-centered $k$-point grids. The same $k$-point grids were employed for both GW and BSE calculations. The cutoff for the screened Coulomb interaction in both GW and BSE corresponds to the dielectric cutoff.}
    \label{tab:gw_bse_paramenters}
\end{table}

Finally, electronic representations in the form of $\mathcal{D}$-matrices~(Eqs.~\eqref{eq:dmats} and \eqref{eq:dmats_trev}) were computed using {\tt LetzElPhC}~\cite{Nalabothula2025May}. \replytoreferee{Python scripts used to construct the representation matrices for excitons, perform symmetry analysis, and compute Raman and phonon-assisted luminescence intensities are now integrated into the \texttt{YamboPy}~\cite{yambopy} project.}

\twocolumngrid
\bibliography{bibilo}
\end{document}